\def\arxiv{}
\newif\iflinenums
\linenumsfalse
\def\docopts{manuscript}
\def\docclass{aastex}
\def\figwidth{0.9\textwidth}
\def\tabletype{deluxetable}

\def\turnpackage{pdflscape}

\ifdefined\aastex
  \def\docopts{preprint}
  \def\docclass{aastex}
  \def\figwidth{0.9\textwidth}
\fi

\ifdefined\arxiv
  \def\docopts{iop,revtex4-1}
  \def\docclass{emulateapj}
  \def\figwidth{\columnwidth}
  \def\tabletype{deluxetable*}

  \def\turnpackage{}
\fi

\ifdefined\lat
  \linenumstrue
\fi

\documentclass[\docopts]{\docclass}

\pdfoutput=1

\usepackage{aas_macros}
\usepackage{xspace}
\usepackage{listings}
\usepackage{lineno}
\usepackage{\turnpackage}

\linenumsfalse
\iflinenums
  \linenumbers
\fi

\usepackage[dvipsnames,svgnames]{xcolor}
\usepackage{graphicx}
\usepackage{afterpage}
\usepackage{natbib}
\usepackage[bookmarks=false]{hyperref}
\usepackage{amsmath}
\usepackage{multirow}
\hypersetup{
  colorlinks      = true,
  linkcolor       = {black},
  linkbordercolor = {white},
  citecolor       = {black},
  citebordercolor = {white},
  urlcolor        = {blue},
  urlbordercolor  = {white},
}

\providecommand\fdg{\mbox{$.\!\!^\circ$}}
\providecommand\facility[1]{facility: #1}

\newcommand{\nobjsDES}{{eight}\xspace}

\newcommand{\ntotal}{{15}\xspace}

\newcommand{\gammaRays}{{$\gamma$ rays}\xspace}
\newcommand{\gammaRayHyph}{{$\gamma$-ray}\xspace}

\newcommand{\kimII}{{Kim~2}\xspace}

\newcommand{\booIII}{{Bo{\"o}tes~III}\xspace}

\newcommand{\wilI}{{Willman~1}\xspace}

\newcommand{\draII}{{Draco~II}\xspace}

\newcommand{\retII}{{Reticulum~II}\xspace}

\newcommand{\tucII}{{Tucana~II}\xspace}

\newcommand{\eriIII}{{Eridanus~III}\xspace}

\newcommand{\indI}{{DES\,J2038\allowbreak$-$4610}\xspace}

\newcommand{\tucIII}{{Tucana~III}\xspace}
\newcommand{\tucIV}{{Tucana~IV}\xspace}
\newcommand{\tucV}{{Tucana~V}\xspace}

\newcommand{\cetII}{{Cetus~II}\xspace}
\newcommand{\indII}{{Indus~II}\xspace}

\newcommand{\irf}[1]{\texttt{#1}\xspace}
\newcommand{\stools}{\emph{ScienceTools}\xspace}
\newcommand{\stool}{\emph{ScienceTool}\xspace}

\newcommand{\ie}{i.e.\xspace}%\def\ie{i.e.}
\newcommand{\eg}{e.g.\xspace}%\def\eg{e.g.}
\newcommand{\FIXME}[1]{}
\newcommand{\CHECK}[1]{{#1}}
\newcommand{\NEW}[1]{{#1}}

\newcommand*\ruleline[1]{\par\noindent\raisebox{1ex}{\makebox[0.97\linewidth]{\hrulefill\quad\raisebox{-.6ex}{#1}\quad\hrulefill}}}

\mathchardef\mhyphen="2D

\newcommand{\vect}[1]{\boldsymbol{#1}}
\newcommand{\roughly}{\ensuremath{ {\sim}\,} }

\newcommand{\unit}[1]{\ensuremath{\mathrm{\,#1}}\xspace}
\newcommand{\dex}{\unit{dex}}
\newcommand{\MeV}{\unit{MeV}}
\newcommand{\GeV}{\unit{GeV}}
\newcommand{\TeV}{\unit{TeV}}
\newcommand{\degree}{\ensuremath{{}^{\circ}}\xspace}

\newcommand{\cm}{\unit{cm}}
\newcommand{\km}{\unit{km}}
\newcommand{\pc}{\unit{pc}}
\newcommand{\kpc}{\unit{kpc}}
\newcommand{\second}{\unit{s}}

\newcommand{\Gyr}{\unit{Gyr}\xspace}
\newcommand{\kms}{{\km \second^{-1}}}
\newcommand{\magn}{\unit{mag}}
\newcommand{\asec}{\unit{arcsec}}

\newcommand{\secref}[1]{Section~\ref{sec:#1}}

\newcommand{\tabref}[1]{Table~\ref{tab:#1}}

\newcommand{\figref}[1]{Figure~\ref{fig:#1}}

\newcommand{\Fermi}{\textit{Fermi}\xspace}

\newcommand{\TS}{\ensuremath{\mathrm{TS}}\xspace}

\newcommand{\pvalue}{\textit{p}-value\xspace}
\newcommand{\pvalues}{\textit{p}-values\xspace}
\newcommand{\pval}{\ensuremath{p}\xspace}
\newcommand{\plocal}{\ensuremath{\pval_{\rm local}}\xspace}
\newcommand{\pobject}{\ensuremath{\pval_{\rm target}}\xspace}
\newcommand{\ptarget}{\ensuremath{\pval_{\rm target}}\xspace}
\newcommand{\psample}{\ensuremath{\pval_{\rm sample}}\xspace}
\newcommand{\pglobal}{\ensuremath{\pval_{\rm global}}\xspace}

\newcommand{\code}[1]{\lstinline!#1!\xspace}

\newcommand{\DM}{\ensuremath{\mathrm{DM}}}

\newcommand{\sigmav}{\ensuremath{\langle \sigma v \rangle}\xspace}

\newcommand{\bbbar}{\ensuremath{b \bar b}\xspace}

\newcommand{\tautau}{\ensuremath{\tau^{+}\tau^{-}}\xspace}

\newcommand{\Jfactor}{\ensuremath{\mathrm{J\mhyphen factor}}\xspace}
\newcommand{\Jfactors}{\ensuremath{\mathrm{J\mhyphen factors}}\xspace}

\newlength{\twocolfigwidth}
\newlength{\onecolfigwidth}
\newlength{\onecolfigwidths}
\graphicspath{{./}{../figures/}}

\begin{document}

\title{Searching for Dark Matter Annihilation in Recently Discovered Milky Way Satellites with Fermi-LAT}
%\input{authors.tex}
% WARNING: Hacking umlaut escape sequence
% Author list file generated with: authlist.py 0.2.0 
% authlist.py -j emulateapj DES-2015-0153_author_list.csv 
\def\andname{}

\author{
A.~Albert\altaffilmark{1,$\dagger$},
B.~Anderson\altaffilmark{2,3},
K.~Bechtol\altaffilmark{4,$\dagger$},
A.~Drlica-Wagner\altaffilmark{5,$\dagger$},
M.~Meyer\altaffilmark{2,3},
M.~S\'anchez-Conde\altaffilmark{3,2},
L.~Strigari\altaffilmark{6,$\dagger$},
M.~Wood\altaffilmark{1,$\dagger$},
T. M. C.~Abbott\altaffilmark{7},
F.~B.~Abdalla\altaffilmark{8,9},
A.~Benoit-L{\'e}vy\altaffilmark{10,8,11},
G.~M.~Bernstein\altaffilmark{12},
R.~A.~Bernstein\altaffilmark{13},
E.~Bertin\altaffilmark{10,11},
D.~Brooks\altaffilmark{8},
D.~L.~Burke\altaffilmark{14,15},
A.~Carnero~Rosell\altaffilmark{16,17},
M.~Carrasco~Kind\altaffilmark{18,19},
J.~Carretero\altaffilmark{20,21},
M.~Crocce\altaffilmark{20},
C.~E.~Cunha\altaffilmark{14},
C.~B.~D'Andrea\altaffilmark{22,23},
L.~N.~da Costa\altaffilmark{16,17},
S.~Desai\altaffilmark{24,25},
H.~T.~Diehl\altaffilmark{5},
J.~P.~Dietrich\altaffilmark{24,25},
P.~Doel\altaffilmark{8},
T.~F.~Eifler\altaffilmark{12,26},
A.~E.~Evrard\altaffilmark{27,28},
A.~Fausti Neto\altaffilmark{16},
D.~A.~Finley\altaffilmark{5},
B.~Flaugher\altaffilmark{5},
P.~Fosalba\altaffilmark{20},
J.~Frieman\altaffilmark{5,29},
D.~W.~Gerdes\altaffilmark{28},
D.~A.~Goldstein\altaffilmark{30,31},
D.~Gruen\altaffilmark{14,15},
R.~A.~Gruendl\altaffilmark{18,19},
K.~Honscheid\altaffilmark{32,33},
D.~J.~James\altaffilmark{7},
S.~Kent\altaffilmark{5},
K.~Kuehn\altaffilmark{34},
N.~Kuropatkin\altaffilmark{5},
O.~Lahav\altaffilmark{8},
T.~S.~Li\altaffilmark{6},
M.~A.~G.~Maia\altaffilmark{16,17},
M.~March\altaffilmark{12},
J.~L.~Marshall\altaffilmark{6},
P.~Martini\altaffilmark{32,35},
C.~J.~Miller\altaffilmark{27,28},
R.~Miquel\altaffilmark{36,21},
E.~Neilsen\altaffilmark{5},
B.~Nord\altaffilmark{5},
R.~Ogando\altaffilmark{16,17},
A.~A.~Plazas\altaffilmark{26},
K.~Reil\altaffilmark{15},
A.~K.~Romer\altaffilmark{37},
E.~S.~Rykoff\altaffilmark{14,15},
E.~Sanchez\altaffilmark{38},
B.~Santiago\altaffilmark{39,16},
M.~Schubnell\altaffilmark{28},
I.~Sevilla-Noarbe\altaffilmark{38,18},
R.~C.~Smith\altaffilmark{7},
M.~Soares-Santos\altaffilmark{5},
F.~Sobreira\altaffilmark{16},
E.~Suchyta\altaffilmark{12},
M.~E.~C.~Swanson\altaffilmark{19},
G.~Tarle\altaffilmark{28},
V.~Vikram\altaffilmark{41},
A.~R.~Walker\altaffilmark{7},
R.~H.~Wechsler\altaffilmark{42,14,15}
\\ \vspace{0.2cm} (The Fermi-LAT and DES Collaborations) \\
}

\affil{$^{\dagger}$ Corresponding authors: A.~Albert, amalbert@lanl.gov; K.~Bechtol, keith.bechtol@icecube.wisc.edu; A.~Drlica-Wagner, kadrlica@fnal.gov; L.~Strigari, strigari@physics.tamu.edu; M.~Wood, mdwood@slac.stanford.edu.} 
\affil{$^{1}$ Los Alamos National Laboratory, Los Alamos, NM 87545, USA}
\affil{$^{2}$ Department of Physics, Stockholm University, AlbaNova, SE-106 91 Stockholm, Sweden}
\affil{$^{3}$ The Oskar Klein Centre for Cosmoparticle Physics, AlbaNova, SE-106 91 Stockholm, Sweden}
\affil{$^{4}$ Dept.  of  Physics  and  Wisconsin  IceCube  Particle  Astrophysics  Center, University  of  Wisconsin, Madison,  WI  53706, USA}
\affil{$^{5}$ Fermi National Accelerator Laboratory, P. O. Box 500, Batavia, IL 60510, USA}
\affil{$^{6}$ George P. and Cynthia Woods Mitchell Institute for Fundamental Physics and Astronomy, and Department of Physics and Astronomy, Texas A\&M University, College Station, TX 77843,  USA}
\affil{$^{7}$ Cerro Tololo Inter-American Observatory, National Optical Astronomy Observatory, Casilla 603, La Serena, Chile}
\affil{$^{8}$ Department of Physics \& Astronomy, University College London, Gower Street, London, WC1E 6BT, UK}
\affil{$^{9}$ Department of Physics and Electronics, Rhodes University, PO Box 94, Grahamstown, 6140, South Africa}
\affil{$^{10}$ CNRS, UMR 7095, Institut d'Astrophysique de Paris, F-75014, Paris, France}
\affil{$^{11}$ Sorbonne Universit\'es, UPMC Univ Paris 06, UMR 7095, Institut d'Astrophysique de Paris, F-75014, Paris, France}
\affil{$^{12}$ Department of Physics and Astronomy, University of Pennsylvania, Philadelphia, PA 19104, USA}
\affil{$^{13}$ Carnegie Observatories, 813 Santa Barbara St., Pasadena, CA 91101, USA}
\affil{$^{14}$ Kavli Institute for Particle Astrophysics \& Cosmology, P. O. Box 2450, Stanford University, Stanford, CA 94305, USA}
\affil{$^{15}$ SLAC National Accelerator Laboratory, Menlo Park, CA 94025, USA}
\affil{$^{16}$ Laborat\'orio Interinstitucional de e-Astronomia - LIneA, Rua Gal. Jos\'e Cristino 77, Rio de Janeiro, RJ - 20921-400, Brazil}
\affil{$^{17}$ Observat\'orio Nacional, Rua Gal. Jos\'e Cristino 77, Rio de Janeiro, RJ - 20921-400, Brazil}
\affil{$^{18}$ Department of Astronomy, University of Illinois, 1002 W. Green Street, Urbana, IL 61801, USA}
\affil{$^{19}$ National Center for Supercomputing Applications, 1205 West Clark St., Urbana, IL 61801, USA}
\affil{$^{20}$ Institut de Ci\`encies de l'Espai, IEEC-CSIC, Campus UAB, Carrer de Can Magrans, s/n,  08193 Bellaterra, Barcelona, Spain}
\affil{$^{21}$ Institut de F\'{\i}sica d'Altes Energies (IFAE), The Barcelona Institute of Science and Technology, Campus UAB, 08193 Bellaterra (Barcelona) Spain}
\affil{$^{22}$ Institute of Cosmology \& Gravitation, University of Portsmouth, Portsmouth, PO1 3FX, UK}
\affil{$^{23}$ School of Physics and Astronomy, University of Southampton,  Southampton, SO17 1BJ, UK}
\affil{$^{24}$ Excellence Cluster Universe, Boltzmannstr.\ 2, 85748 Garching, Germany}
\affil{$^{25}$ Faculty of Physics, Ludwig-Maximilians-Universit\"at, Scheinerstr. 1, 81679 Munich, Germany}
\affil{$^{26}$ Jet Propulsion Laboratory, California Institute of Technology, 4800 Oak Grove Dr., Pasadena, CA 91109, USA}
\affil{$^{27}$ Department of Astronomy, University of Michigan, Ann Arbor, MI 48109, USA}
\affil{$^{28}$ Department of Physics, University of Michigan, Ann Arbor, MI 48109, USA}
\affil{$^{29}$ Kavli Institute for Cosmological Physics, University of Chicago, Chicago, IL 60637, USA}
\affil{$^{30}$ Department of Astronomy, University of California, Berkeley,  501 Campbell Hall, Berkeley, CA 94720, USA}
\affil{$^{31}$ Lawrence Berkeley National Laboratory, 1 Cyclotron Road, Berkeley, CA 94720, USA}
\affil{$^{32}$ Center for Cosmology and Astro-Particle Physics, The Ohio State University, Columbus, OH 43210, USA}
\affil{$^{33}$ Department of Physics, The Ohio State University, Columbus, OH 43210, USA}
\affil{$^{34}$ Australian Astronomical Observatory, North Ryde, NSW 2113, Australia}
\affil{$^{35}$ Department of Astronomy, The Ohio State University, Columbus, OH 43210, USA}
\affil{$^{36}$ Instituci\'o Catalana de Recerca i Estudis Avan\c{c}ats, E-08010 Barcelona, Spain}
\affil{$^{37}$ Department of Physics and Astronomy, Pevensey Building, University of Sussex, Brighton, BN1 9QH, UK}
\affil{$^{38}$ Centro de Investigaciones Energ\'eticas, Medioambientales y Tecnol\'ogicas (CIEMAT), Madrid, Spain}
\affil{$^{39}$ Instituto de F\'\i sica, UFRGS, Caixa Postal 15051, Porto Alegre, RS - 91501-970, Brazil}
\affil{$^{40}$ Computer Science and Mathematics Division, Oak Ridge National Laboratory, Oak Ridge, TN 37831, USA}
\affil{$^{41}$ Argonne National Laboratory, 9700 South Cass Avenue, Lemont, IL 60439, USA}
\affil{$^{42}$ Department of Physics, Stanford University, 382 Via Pueblo Mall, Stanford, CA 94305, USA}

%TC:break Abstract
\begin{abstract}
We search for excess \gammaRayHyph emission coincident with the positions of confirmed and candidate Milky Way satellite galaxies using 6 years of data from the \Fermi Large Area Telescope (LAT).
% Previously Known (22) : 27 satellites - LMC - SMC - Sgr - CMa - Seg 2
% New Confirmed (6): Dra II + Hor I + Hya II + Ret II + Tri II + Tuc II
% New Systems: Kim 2 + Peg III + 7 Y1 + Gru I + Hya II + Hor II + 8 Y2 + Sgr II + Dra II + Tri II
Our sample of \CHECK{45} stellar systems includes \CHECK{28} kinematically confirmed dark-matter-dominated dwarf spheroidal galaxies (dSphs) and \CHECK{17} recently discovered systems that have photometric characteristics consistent with the population of known dSphs.
For each of these targets, the relative predicted \gammaRayHyph flux due to dark matter annihilation is taken from kinematic analysis if available, and estimated from a distance-based scaling relation otherwise, assuming that the stellar systems are dark-matter-dominated dSphs.
LAT data coincident with four of the newly discovered targets show a slight preference (each \CHECK{$\roughly 2 \sigma$} local) for \gammaRayHyph emission in excess of the background.
However, the ensemble of derived \gammaRayHyph flux upper limits for individual targets is consistent with the expectation from analyzing random blank-sky regions, and a combined analysis of the population of stellar systems yields no globally significant excess (global significance \CHECK{$<1\sigma$}).
Our analysis has increased sensitivity compared to the analysis of 15 confirmed dSphs by \citet{Ackermann:2015zua}. 
The observed constraints on the dark matter annihilation cross section are statistically consistent with the background expectation, improving by a factor of $\roughly 2$ for large dark matter masses ($m_{{\rm DM}, \bbbar} \gtrsim 1\TeV$ and $m_{{\rm DM}, \tautau} \gtrsim 70 \GeV$) and weakening by a factor of $\roughly 1.5$ at lower masses relative to previously observed limits.


\end{abstract}

\maketitle

\section{Introduction}\label{sec:intro}

Astrophysical evidence suggests that non-baryonic cold dark matter (DM) constitutes $\roughly 84\%$ of the matter density of the Universe~\citep{Ade:2015xua}.
Many particle DM candidates, such as weakly interacting massive particles (WIMPs), are predicted to annihilate or decay into energetic Standard Model particles \citep[\eg,][]{Bertone:2004pz,Feng:2010gw}.
Depending on the DM particle mass and annihilation cross section or decay rate, these interactions may produce \gammaRays detectable by instruments such as the \Fermi Large Area Telescope (LAT), which is sensitive to \gammaRays in the range from $20\MeV$ to ${>}300\GeV$ \citep{Atwood:2009ez}.
Milky Way dwarf spheroidal satellite galaxies (dSphs) are excellent targets to search for \gammaRays produced from DM annihilation due to their proximity, their large DM density, and the absence of observational evidence for non-thermal astrophysical processes that produce \gammaRays \citep[\eg,][]{Evans:2003sc,Baltz:2008wd}.

The expected \gammaRayHyph flux from DM annihilation is
\begin{equation}
\begin{aligned}
       \phi(\Delta\Omega,E_{\min},E_{\max}) =
    & \underbrace{ \frac{1}{4\pi} \frac{\sigmav}{2m_{\DM}^{2}}\int^{E_{\max}}_{E_{\min}}\frac{\text{d}N_{\gamma}}{\text{d}E_{\gamma}}\text{d}E_{\gamma}}_{\rm particle~physics}\\
    &    \times
    \underbrace{\vphantom{\int_{E_{\min}}} \int_{\Delta\Omega}\int_{\rm l.o.s.}\rho_{\DM}^{2}(\vect{r}(l))\text{d}l\text{d}\Omega }_{\rm \Jfactor}\,,
\end{aligned}
\label{eqn:annihilation}
\end{equation}
\noindent where \sigmav is the velocity-averaged DM annihilation cross section, $m_{\mathrm{DM}}$ is the DM particle mass, and $\frac{dN_\gamma}{dE}$ is the differential \gammaRayHyph\ photon counts spectrum summed over all final states.
The ``\Jfactor'' is the square of the DM density ($\rho$) as a function of position $\vect{l}$ integrated along the light-of-sight (l.o.s.) in the region of interest (ROI), and $\Delta \Omega$ denotes the solid angle over which the \Jfactor is calculated \citep{Gondolo:2004sc}.

The \Jfactors of dSphs can be inferred from the measured velocities of their member stars \citep[\eg,][]{Simon:2007dq,Walker:2008ax}.
While the \Jfactors of individual dSphs are several orders of magnitude smaller than that of the Galactic center, observations of individual dSphs can be combined to increase the sensitivity to a DM annihilation signal while simultaneously reducing the impact of systematic uncertainties for individual dSphs.
In addition, observations of the dSphs provide an important independent test of DM interpretations of the \gammaRayHyph excess associated with the Galactic center \citep[GCE;][]{Gordon:2013vta,Daylan:2014rsa,Abazajian:2014fta,Calore:2014xka,Ajello:2015kwa}.

\NEW{
Multiple groups have searched for excess \gammaRays associated with dSphs using LAT data and have reported constraints on DM annihilation that are competitive with other DM targets such as the Galactic center \citep[\eg,][]{Abdo:2010ex,Ackermann:2011wa,GeringerSameth:2011iw,Mazziotta:2012ux,Ackermann:2013yva,Geringer-Sameth:2014qqa,Geringer-Sameth:2015lua,Hooper:2015ula,Li:2015kag}.
For example, the combined likelihood analysis of 15 dSphs with 6 years of LAT \irf{Pass 8} data by \citet{Ackermann:2015zua} excludes DM particles with masses ${\lesssim} 100 \GeV$ annihilating with the canonical thermal relic cross section via quark or $\tau$-lepton channels.
That work used only dSphs with spectoscopically determined \Jfactors.
}

In 2015, a combination of on-going wide-field optical imaging surveys and a re-analysis of Sloan Digital Sky Survey \citep[SDSS;][]{York:2000gk} data revealed more than \CHECK{20} new satellite systems~\citep{Bechtol:2015wya,Koposov:2015cua,Laevens:2015kla,2015ApJ...802L..18L,2015ApJ...804L..44K,2015ApJ...804L...5M,2015ApJ...808L..39K,Drlica-Wagner:2015b}.
The photometric characteristics of these new Milky Way satellites are consistent with previously known dSphs, but are referred to as ``dSph candidates'' until their DM content is spectroscopically confirmed.

If the newly discovered systems are confirmed as DM-dominated dSphs, they represent important new targets in the search for \gammaRays from DM annihilation.
This paper follows on the work of \citet[][hereafter DW15]{Drlica-Wagner:2015xua}, who analyzed satellites discovered in the first year (Y1) of the Dark Energy Survey \citep[DES;][]{Abbott:2005bi}.
Here, we perform a \NEW{comprehensive} \Fermi-LAT \gammaRayHyph analysis of all confirmed and candidate dSphs.
This target sample includes \ntotal additional dSph candidates found in year two (Y2) of DES and other surveys.
In total, our sample comprises 45 confirmed and candidate dSphs (\secref{targets}).
We find slight (\CHECK{$\roughly 2 \sigma$} local) excesses of \gammaRays coincident with four of the new targets (\secref{LAT}).
Spectroscopic observations are needed to measure the dynamical masses and associated \Jfactors of the new systems.
For recently discovered dSph candidates that lack spectroscopic observations, we use a simple scaling relation to predict \Jfactors based on photometric data alone (\secref{jfactors}).
In \secref{darkmatter} we perform a combined analysis of the population of confirmed and candidate dSphs and find no globally significant excess associated with the ensemble of targets.
We therefore present constraints on the DM annihilation cross section derived from the population of confirmed and candidate dSphs.
In \secref{conclusions} we summarize our findings and conclude.

\section{Targets}
\label{sec:targets}

In 2015, wide-field optical imaging surveys enabled the discovery of more than 20 new Milky Way satellites having morphological characteristics similar to the known DM-dominated dSphs.
Each of these satellites was identified as a statistically significant arcminute-scale overdensity of resolved stars consistent with an old (${>} 10 \Gyr$) and metal-poor ($Z \sim 0.0002$) simple stellar population.
The basic structural characteristics of each stellar system (\eg, center position, heliocentric distance, and spatial extension) were inferred by fitting the spatial and color-magnitude distributions of probable member stars.

The majority of the recently announced Milky Way satellites were discovered in DES data collected with the Dark Energy Camera \citep[DECam;][]{Flaugher:2015}.
Searches of the DES Y1 data by both the DES Collaboration and other groups led to the discovery of nine dSph candidates \citep{Bechtol:2015wya,Koposov:2015cua,2015ApJ...808L..39K}.
\citet{Drlica-Wagner:2015b} subsequently reported \nobjsDES additional dSph candidates found in DES Y2 data.
The compact stellar systems Kim~2 \citep{Kim:2015a}\footnote{Kim~2 was also identified as \indI /Indus~I by \citet{Bechtol:2015wya} and \citet{Koposov:2015cua} slightly after its original discovery by \citet{Kim:2015a}.} and DES~1 \citep{Luque:2015a} are also present in the DES data; however, they have photometric properties that are more similar to low-luminosity outer-halo star clusters than to dSphs.

In addition to the objects found in DES data, several systems have recently been discovered in other surveys.
Pegasus~III  was detected in archival SDSS data \citep{2015ApJ...804L..44K} and later confirmed as a stellar overdensity with DECam.
Hydra~II was found serendipitously in DECam data taken for the Survey of the MAgellanic Stellar History~\citep{2015ApJ...804L...5M}.
Three additional dSph candidates were discovered in the Pan-STARRS~1 $3\pi$ Survey: Triangulum~II, Draco~II, and Sagittarius~II \citep{2015ApJ...802L..18L,Laevens:2015kla}.
We note that several other systems have been discovered using Pan-STARRS~1, but due to their small sizes and/or measured kinematics, they are classified as globular clusters and are not considered in this work~\citep{Laevens:2014fia,Laevens:2015kla,2015arXiv150601021K}.

Thus far, six recently discovered systems have measured kinematics consistent with being DM-dominated dSphs: Reticulum~II \citep{Simon:2015fdw,Walker:2015,Koposov:2015b}, Horologium~I \citep{Koposov:2015b}, Hydra~II \citep{2015arXiv150601021K}, Draco~II \citep{2015arXiv151001326M}, Triangulum~II \citep{Kirby:2015bxa,2015arXiv151004433M}, and Tucana~II \citep{2015arXiv151106296W}.
\Jfactors have been derived for Reticulum~II \citep{Simon:2015fdw,Bonnivard:2015tta} and Tucana~II \citep{2015arXiv151106296W} from these kinematic data.

\newcommand{\tblcaption}{Confirmed and Candidate Dwarf Galaxies}
\newcommand{\tblcomments}{\label{tab:targets}
\scriptsize Milky Way satellite systems consistent with being dSphs. Horizontal lines divide systems that have been kinematically determined to be DM dominated (top), systems with photometry consistent with being dSphs (middle), and systems with small physical sizes populating an ambiguous region of the size-luminosity plane between dSphs and globular clusters (bottom). Columns represent (1) name of stellar system (2) Galactic coordinates (3) heliocentric distance (4) azimuthally averaged half-light radius (5) absolute visual magnitude (6) measured J-factor derived from stellar kinematics by \citet{Geringer-Sameth:2014yza}; \NEW{\retII value taken from \citet{Simon:2015fdw}} (7) predicted \Jfactor from Equation~\ref{eqn:jpred} (8) composite sample membership (see \secref{darkmatter}): C=conservative, N=nominal, I=inclusive.  Targets used in the combined limits from \citet{Ackermann:2015zua} are marked with asterisks.
}

\begin{\tabletype}{l ccccccc }
\tablewidth{0pt}
\tabletypesize{\tiny}
\tablecaption{ \tblcaption }
\tablehead{
(1) & (2) & (3) & (4) & (5) & (6) & (7) & (8)\\
Name & $l,b$ & Distance & $r_{1/2}$ & $M_{V}$ & $\log_{10}(J_{\rm meas})$ & $\log_{10}(J_{\rm pred})$ & Sample \\
 & (deg, deg) & (kpc) & (pc) & (mag) & $\log_{10}(\GeV^2 \cm^{-5})$ & $\log_{10}(\GeV^2 \cm^{-5})$ & 
}
\startdata
\multicolumn{8}{c}{\ruleline{Kinematically Confirmed Galaxies}}\\
Bo{\"o}tes I*                & 358.08, 69.62  & 66  & 189  & -6.3 & $18.2 \pm 0.4$ & 18.5           & I,N,C \\
Bo{\"o}tes II                & 353.69, 68.87  & 42  & 46   & -2.7 & ...            & 18.9           & I,N,C \\
Bo{\"o}tes III               & 35.41, 75.35   & 47  & ...  & -5.8 & ...            & 18.8           & I,N \\
Canes Venatici I             & 74.31, 79.82   & 218 & 441  & -8.6 & $17.4 \pm 0.3$ & 17.4           & I,N,C \\
Canes Venatici II*           & 113.58, 82.70  & 160 & 52   & -4.9 & $17.6 \pm 0.4$ & 17.7           & I,N,C \\
Carina*                      & 260.11, -22.22 & 105 & 205  & -9.1 & $17.9 \pm 0.1$ & 18.1           & I,N,C \\
Coma Berenices*              & 241.89, 83.61  & 44  & 60   & -4.1 & $19.0 \pm 0.4$ & 18.8           & I,N,C \\
Draco*                       & 86.37, 34.72   & 76  & 184  & -8.8 & $18.8 \pm 0.1$ & 18.3           & I,N,C \\
Draco II                     & 98.29, 42.88   & 24  & 16   & -2.9 & ...            & 19.3           & I,N,C \\
Fornax*                      & 237.10, -65.65 & 147 & 594  & -13.4 & $17.8 \pm 0.1$ & 17.8           & I,N,C \\
Hercules*                    & 28.73, 36.87   & 132 & 187  & -6.6 & $16.9 \pm 0.7$ & 17.9           & I,N,C \\
Horologium I                 & 271.38, -54.74 & 87  & 61   & -3.5 & ...            & 18.2           & I,N,C \\
Hydra II                     & 295.62, 30.46  & 134 & 66   & -4.8 & ...            & 17.8           & I,N,C \\
Leo I                        & 225.99, 49.11  & 254 & 223  & -12.0 & $17.8 \pm 0.2$ & 17.3           & I,N,C \\
Leo II*                      & 220.17, 67.23  & 233 & 164  & -9.8 & $18.0 \pm 0.2$ & 17.4           & I,N,C \\
Leo IV*                      & 265.44, 56.51  & 154 & 147  & -5.8 & $16.3 \pm 1.4$ & 17.7           & I,N,C \\
Leo V                        & 261.86, 58.54  & 178 & 95   & -5.2 & $16.4 \pm 0.9$ & 17.6           & I,N,C \\
Pisces II                    & 79.21, -47.11  & 182 & 45   & -5.0 & ...            & 17.6           & I,N,C \\
Reticulum II                 & 266.30, -49.74 & 32  & 35   & -3.6 & $18.9 \pm 0.6$ & 19.1           & I,N,C \\
Sculptor*                    & 287.53, -83.16 & 86  & 233  & -11.1 & $18.5 \pm 0.1$ & 18.2           & I,N,C \\
Segue 1*                     & 220.48, 50.43  & 23  & 21   & -1.5 & $19.4 \pm 0.3$ & 19.4           & I,N,C \\
Sextans*                     & 243.50, 42.27  & 86  & 561  & -9.3 & $17.5 \pm 0.2$ & 18.2           & I,N,C \\
Triangulum II                & 140.90, -23.82 & 30  & 30   & -1.8 & ...            & 19.1           & I,N,C \\
Tucana II                    & 328.04, -52.35 & 58  & 120  & -3.9 & ...            & 18.6           & I,N,C \\
Ursa Major I                 & 159.43, 54.41  & 97  & 143  & -5.5 & $17.9 \pm 0.5$ & 18.1           & I,N,C \\
Ursa Major II*               & 152.46, 37.44  & 32  & 91   & -4.2 & $19.4 \pm 0.4$ & 19.1           & I,N,C \\
Ursa Minor*                  & 104.97, 44.80  & 76  & 120  & -8.8 & $18.9 \pm 0.2$ & 18.3           & I,N,C \\
Willman 1*                   & 158.58, 56.78  & 38  & 19   & -2.7 & ...            & 18.9           & I,N \\
\multicolumn{8}{c}{\ruleline{Likely Galaxies}}\\
Columba I                    & 231.62, -28.88 & 182 & 101  & -4.5 & ...            & 17.6           & I,N,C \\
Eridanus II                  & 249.78, -51.65 & 331 & 156  & -7.4 & ...            & 17.1           & I,N,C \\
Grus I                       & 338.68, -58.25 & 120 & 60   & -3.4 & ...            & 17.9           & I,N,C \\
Grus II                      & 351.14, -51.94 & 53  & 93   & -3.9 & ...            & 18.7           & I,N,C \\
Horologium II                & 262.48, -54.14 & 78  & 33   & -2.6 & ...            & 18.3           & I,N,C \\
Indus II                     & 354.00, -37.40 & 214 & 181  & -4.3 & ...            & 17.4           & I,N,C \\
Pegasus III                  & 69.85, -41.81  & 205 & 57   & -4.1 & ...            & 17.5           & I,N,C \\
Phoenix II                   & 323.69, -59.74 & 96  & 33   & -3.7 & ...            & 18.1           & I,N,C \\
Pictor I                     & 257.29, -40.64 & 126 & 44   & -3.7 & ...            & 17.9           & I,N,C \\
Reticulum III                & 273.88, -45.65 & 92  & 64   & -3.3 & ...            & 18.2           & I,N,C \\
Sagittarius II               & 18.94, -22.90  & 67  & 34   & -5.2 & ...            & 18.4           & I,N,C \\
Tucana III                   & 315.38, -56.18 & 25  & 44   & -2.4 & ...            & 19.3           & I,N \\
Tucana IV                    & 313.29, -55.29 & 48  & 128  & -3.5 & ...            & 18.7           & I,N,C \\
\multicolumn{8}{c}{\ruleline{Ambiguous Systems}}\\
Cetus II                     & 156.47, -78.53 & 30  & 17   & 0.0  & ...            & 19.1           & I \\
Eridanus III                 & 274.95, -59.60 & 96  & 12   & -2.4 & ...            & 18.1           & I \\
Kim 2                        & 347.16, -42.07 & 105 & 12   & -1.5 & ...            & 18.1           & I \\
Tucana V                     & 316.31, -51.89 & 55  & 16   & -1.6 & ...            & 18.6           & I 

\enddata
{\footnotesize \tablecomments{ \tblcomments }}
\end{\tabletype}

\begin{figure}%[Th]
  \begin{centering}
\includegraphics[width=\figwidth]{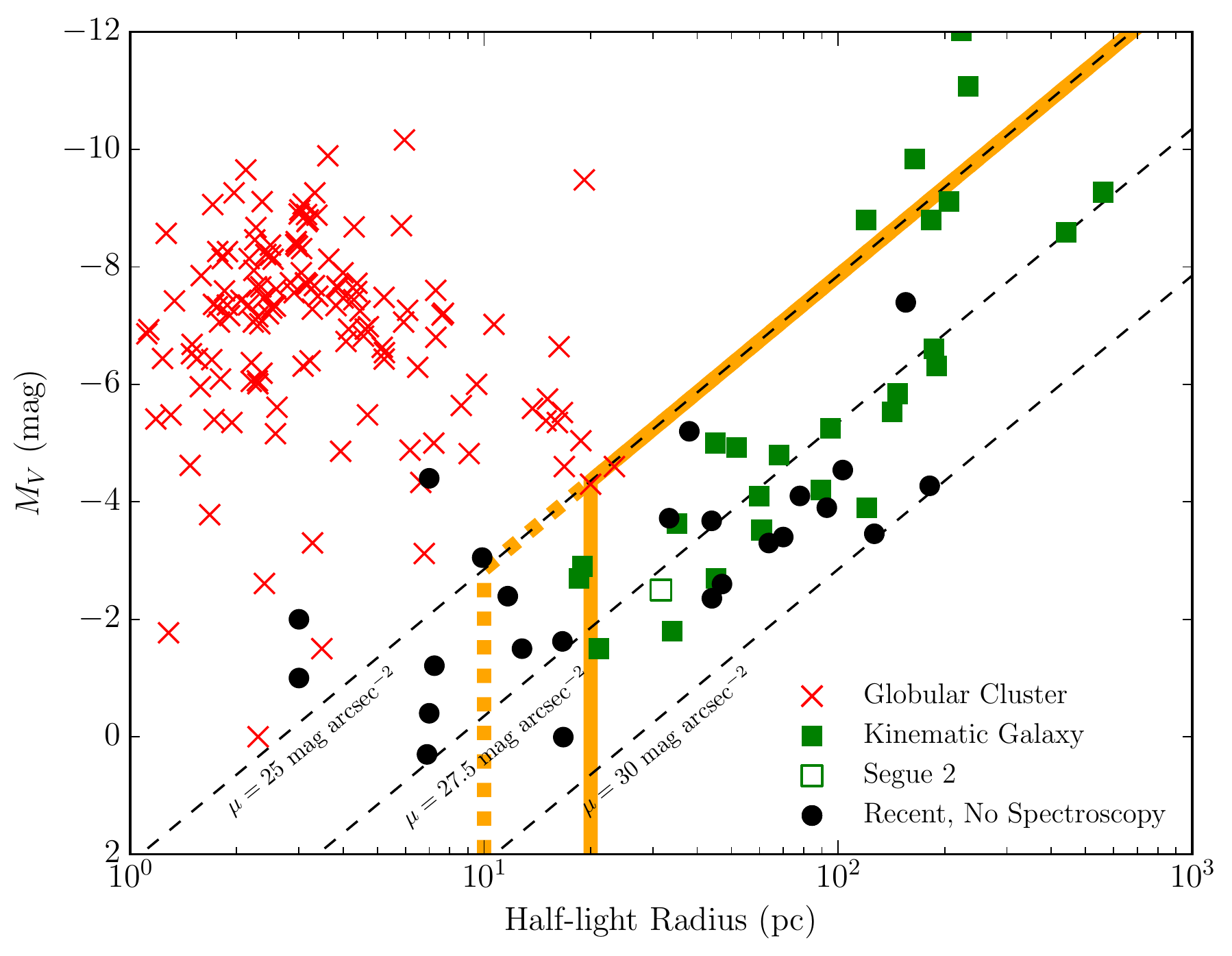}
  \caption{\label{fig:size_luminosity} Absolute visual magnitude ($M_V$) versus physical half-light radius ($r_{1/2}$) for dSphs and globular clusters.
Globular clusters, which do not contain measurable DM within their visible stellar distribution, are marked with red crosses \citep[][2010 edition]{Fadely:2011,Harris:1996}.
Spectroscopically confirmed DM-dominated dSphs are labeled with filled green squares.
Segue~2 (open green square) has the chemical signatures of a dSph, but exhibits a low velocity dispersion \citep{Kirby:2013isa}, and is therefore excluded from our target list.
Milky Way satellites lacking spectroscopic observations are labeled with black filled circles.
Thick orange lines indicate our target sample selection cuts on objects lacking spectroscopic data (see \secref{darkmatter}): nominal (solid; $r_{1/2} > 20 \pc$) and inclusive (dashed; $r_{1/2} > 10 \pc$).
Black dashed lines indicate contours of constant surface brightness ($\mu$).
  }
\end{centering}
\end{figure}

The dSphs are good targets for DM searches because their dynamical and chemical properties suggest the presence of large quantities of DM.
In contrast, globular clusters have mass-to-light ratios of order unity.
Low-luminosity stellar systems cannot be conclusively classified as dSphs or globular clusters without radial velocity measurements.
However, dSphs are generally found to have larger physical half-light radii ($r_{1/2}$) and lower surface brightnesses ($\mu$) than globular clusters (\figref{size_luminosity}).
Therefore, we used the photometric characteristics of the newly discovered systems to select those that are likely to be DM-dominated dSphs when spectroscopic measurements were unavailable.

For stellar systems with $M_V \lesssim -5$, DM-dominated dSphs have $r_{1/2} \gtrsim 100 \pc$, while globular clusters have $r_{1/2} \lesssim 20 \pc$.
For fainter systems, the size distinction becomes less clear.
The most compact kinematically classified dSph is Segue 1~\citep{Geha:2008zr,Simon:2010ek}, which has an azimuthally averaged half-light radius of $21\pc$.
On the other hand, the Palomar~13 globular cluster has a half-light radius of approximately $10\pc$, and does not require DM to explain its measured velocity dispersion~\citep{Bradford:2011aq}.
We note that recent spectroscopy of \draII, which has an azimuthally averaged half-light radius of $16 \pc$, indicates a velocity dispersion $2.9 \pm 2.1 \kms$ and therefore is moderately likely to be DM dominated.~\citep{2015arXiv151001326M}.
We inclusively selected new objects with $r_{1/2} > 10 \pc$ and surface brightnesses $\mu > 25 \magn \asec^{-2}$.

Two confirmed globular clusters (Palomar~14 and Laevens~1) would pass our nominal selection criteria based on their physical size and luminosity ($r_{1/2} \sim 20 \pc$, $M_V \sim -4.5 \magn$).
However, Palomar~14 is kinematically determined to have a mass-to-light ratio near unity \citep{Jordi:2009}, and the relatively large metallicity and low metallicity dispersion of Laevens~1 is more similar to globular clusters \citep{Kirby:2013isa}.
Therefore, we do not include these two systems in our analysis.

In \tabref{targets} we summarize the characteristics of confirmed and candidate dSphs considered in this work.
This table is divided into three sections: (1) systems that are kinematically determined to be DM-dominated dSphs, (2) systems with photometric characteristics consistent with known dSphs, and (3) systems with small physical sizes ($10 \pc < r_{1/2} < 20 \pc$) and ambiguous classifications (see \figref{size_luminosity}).
Due to small stellar samples and/or complicated kinematics, several kinematically confirmed dSphs lack spectroscopically measured \Jfactors.

Several Milky Way satellites are not considered in this analysis.
For instance, the Sagittarius and Canis~Major dSphs are excluded because: (1) they reside at low Galactic latitude ($b = -14\fdg2$ and $b = -8\fdg0$, respectively) where the diffuse Galactic \gammaRayHyph foreground emission presents both statistical and systematic challenges, and (2) they show strong evidence of tidal disruption, making accurate determination of their DM masses difficult \citep{Frinchaboy:2012,Martin:2004}.
In spite of these obstacles, the proximity (26\kpc and 7\kpc, respectively) and large velocity dispersions of these two systems make them promising targets for dedicated individual study.

Finally, we exclude Segue~2 from our target list.
Spectroscopic measurements show that Segue~2 has a large metallicity dispersion characteristic of dSphs, but the upper bound on its velocity dispersion, $\sigma_v <2.2 \kms$, implies a mass-to-light ratio within the half-light radius, $(M/L_V)_{1/2} < 360$ $\mathrm{M}_{\odot} / \mathrm{L}_{\odot}$, lower than that of other comparably luminous dSphs \citep{Kirby:2013isa}.
As shown in \figref{size_luminosity}, Segue~2 is situated within the locus of DM-dominated dSphs according to its photometric properties, and therefore provides a cautionary example of a system that might not follow the scaling relation described in \secref{jfactors}, which assumes a common value for the central DM density of dSphs.

\section{LAT Analysis}
\label{sec:LAT}

\begin{figure*}%[Th]
  \begin{centering}
  \includegraphics[width=\textwidth]{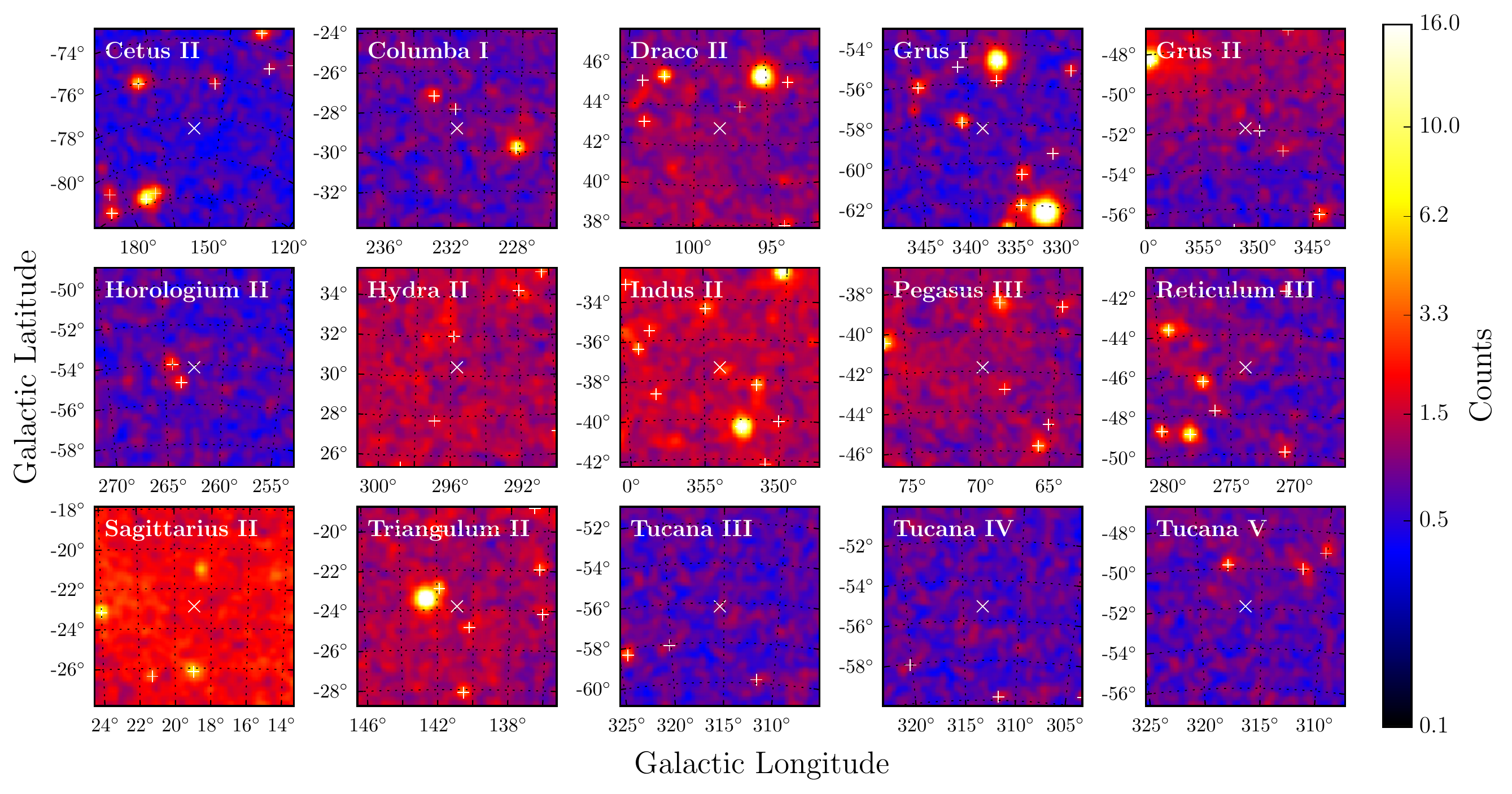}
  \caption{\label{fig:LATCounts}
  Binned \gammaRayHyph counts maps ($E > 1 \GeV$) for $10\degree \times 10 \degree$ ROIs centered on 15 targets that were not analyzed by \citet{Drlica-Wagner:2015xua} or \citet{Ackermann:2015zua}.
  The dSph candidates are indicated with white ``$\times$'' symbols, while 3FGL sources in the ROI are indicated with white ``$+$'' symbols.
  The counts maps are binned in $0\fdg1 \times 0\fdg1$ spatial pixels and smoothed with a $0\fdg25$ Gaussian kernel.
}
\end{centering}
\end{figure*}

We analyzed \gammaRayHyph data coincident with our targets using the same analysis procedure and data set described in \citet{Ackermann:2015zua}.
We briefly review the details of the analysis here for completeness.
Our data set consisted of six years of LAT data (2008 August 4 to 2014 August 5) in the energy range from 500 MeV to 500 GeV passing the \irf{P8R2 SOURCE} event class selections.
We rejected events with zenith angles greater than 100$^{\circ}$ to remove \gammaRays produced by cosmic-ray interactions in the Earth's atmosphere.
Additionally, events from time intervals around bright \gammaRayHyph bursts and solar flares were removed using the same procedure as the third LAT source catalog \citep[3FGL;][]{Ackermann:2015hja}.
To analyze the targets in \tabref{targets}, we used $10 \degree \times 10\degree$ ROIs centered on each target.
In \figref{LATCounts} we show \gammaRayHyph counts maps for 15 systems that were not previously analyzed by DW15 or \citet{Ackermann:2015zua}.
Data reduction was performed using the \Fermi \stools\footnote{\url{http://fermi.gsfc.nasa.gov/ssc/data/analysis/software/}} version 10-01-01 and the \irf{P8R2\_SOURCE\_V6} instrument response functions.\footnote{\url{http://www.slac.stanford.edu/exp/glast/groups/canda/lat_Performance.htm}}

To search for \gammaRayHyph emission coincident with our targets in excess of the local background expectation, we performed a binned maximum-likelihood analysis in 24 logarithmically spaced energy bins and $0\fdg1$ spatial pixels.
Data were partitioned into four point-spread function event types, which were combined in a joint likelihood function when fitting each ROI \citep{Ackermann:2015zua}.
We modeled the Galactic diffuse emission with the standard LAT interstellar emission model ({\it gll\_iem\_v06.fits}) recommended for analysis of the \irf{Pass 8} data.\footnote{\url{http://fermi.gsfc.nasa.gov/ssc/data/access/lat/BackgroundModels.html}}
Additionally, we modeled extragalactic \gammaRayHyph emission and residual charged particle contamination with an isotropic model fit to the \irf{Pass 8} data.
Point sources from the 3FGL catalog within $15^{\circ}$ of the ROI center were included in the background model.
The flux normalizations of the Galactic, isotropic, and 3FGL catalog sources within the $10^{\circ}\times 10^{\circ}$ ROI were fit simultaneously over the broadband energy range from 500\MeV to 500\GeV.
The spectral parameters of all other background components were fixed to their nominal values during the fit.
Following DW15, we enabled the energy dispersion correction in our fits for all components except the Galactic diffuse emission model and the isotropic model.
The flux normalizations of the background sources were insensitive to the inclusion of a putative power-law source at the locations of the targets.
Each ROI was found to be well described by the background model with no significant \NEW{($>3\sigma$)} residuals \NEW{associated with the target locations}.

\begin{figure*}%[Th]
  \begin{centering}
  \includegraphics[width=\textwidth]{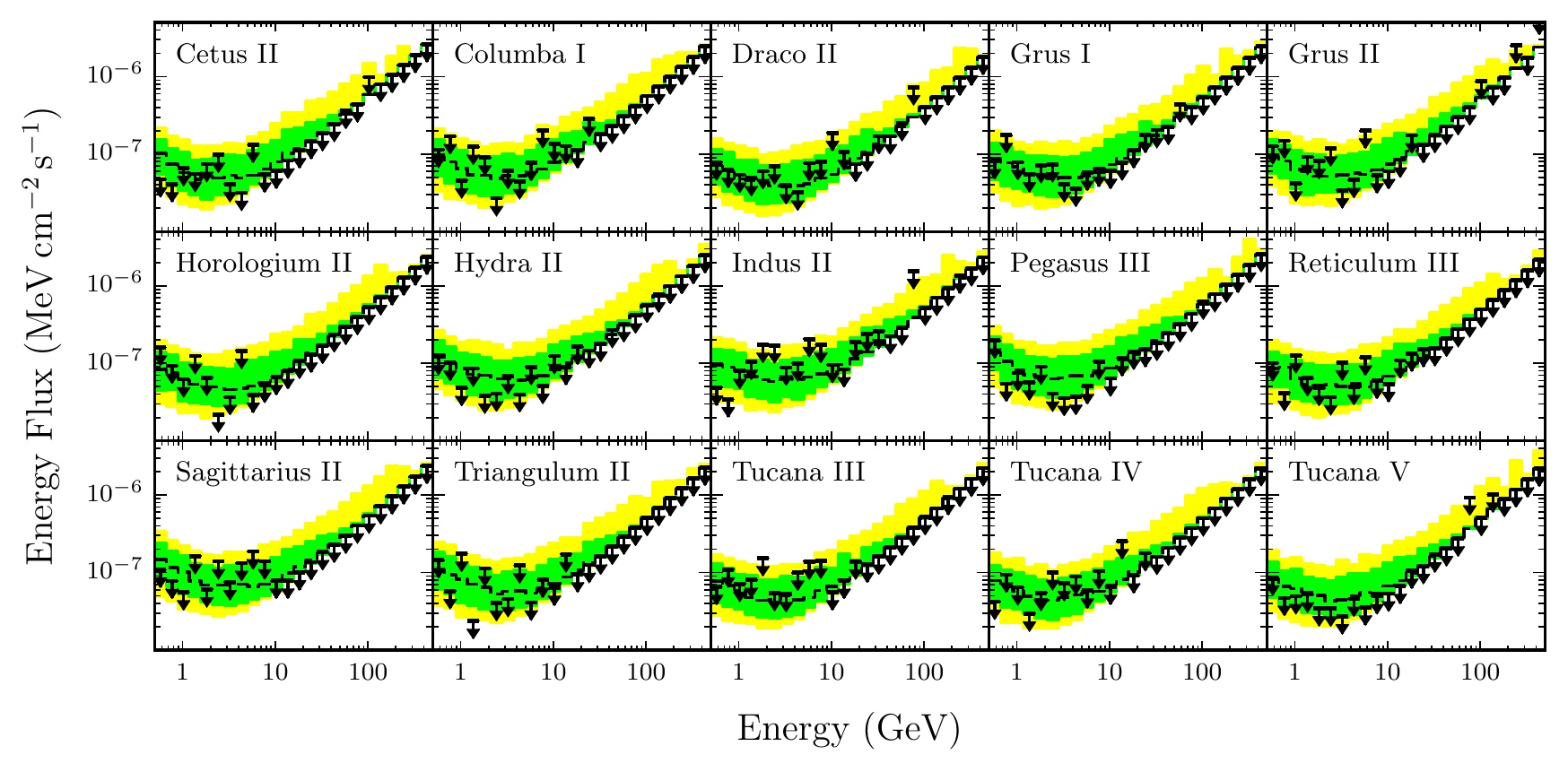}
  \caption{\label{fig:sed_grid}
Bin-by-bin integrated energy-flux upper limits at 95\% confidence level assuming a point-like model for the 15 targets in \figref{LATCounts}.
The median expected sensitivity is shown by the dashed black line while the 68\% and 95\% containment regions are indicated by the green and yellow bands, respectively.
The expected sensitivity and containment regions are derived from \CHECK{300} Monte Carlo simulations of the \gammaRayHyph background in the regions surrounding each respective target.
}
\end{centering}
\end{figure*}

In this analysis we modeled all 45 confirmed and candidate dSphs in \tabref{targets} as point-like sources.
This differs from the analysis of \citet{Ackermann:2015zua} where some targets were modeled as extended sources.
This choice was motivated by a desire to have a consistent analysis across the targets and the fact that the physical sizes of the DM halos surrounding the dSph candidates are not well constrained (we further investigate the impact of this choice in \secref{darkmatter}).
We fit for excess \gammaRayHyph emission above the background associated with each target in each energy bin separately to derive upper limits on the flux that are independent of the choice of spectral model.
Within each bin, we model the putative dSph source with a power-law spectral model ($\mathrm{d}N/\mathrm{d}E$ $\propto E^{-\Gamma}$) with spectral index $\Gamma = 2$ \citep{Ackermann:2013yva,Ackermann:2015zua}.
In \figref{sed_grid} we show the bin-by-bin integrated energy-flux upper limits at 95\% confidence level for 15 dSph candidates not included in DW15 or \citet{Ackermann:2015zua}.
We generate 68\% and 95\% containment bands for the bin-by-bin limits from 300 Monte Carlo simulations of the local \gammaRayHyph background in the region of each dSph using the \Fermi \stool, \code{gtobssim}.
We use simulations to account for local variations in the diffuse \gammaRayHyph background between the individual dSphs \citep{Ackermann:2013yva}.
%rather than the random blank fields

The $10\degree \times 10\degree$ ROIs of several targets overlap, and we investigated possible correlations between the normalization of the putative sources.
The two targets with the smallest angular separation ($\roughly 1\fdg5$) are \tucIII and \tucIV.
We simultaneously fit the normalizations of the Galactic diffuse emission model, \tucIII, and \tucIV in several energy bins, and found the magnitude of the correlation factor between the normalizations of the two dSph candidates to be ${<}0.1$.
The LAT is certainly capable of resolving more closely spaced sources (see Fig. 13 of \citealt{Ackermann:2015hja}), and our result is consistent with that of \citet{Carlson:2014nra}, who studied the correlation between dSph targets and nearby unresolved sources.

To maximize the sensitivity to specific DM spectral models, the Poisson likelihoods from each bin were combined to form broadband likelihoods for different DM annihilation channels and masses.
We tested for excess \gammaRayHyph emission consistent with two representative DM annihilation channels (\ie, \bbbar and \tautau), and scanned a range of DM particle masses in six steps per decade from 2 \GeV to 10 \TeV (when kinematically allowed). \NEW{The spectra were obtained from \code{DMFit} based on \code{PYTHIA 8.165} using the \Fermi \stools \citep{Jeltema:2008hf,Ackermann:2013yva,Ackermann:2015zua}.}
We calculated a test statistic (\TS) for \gammaRayHyph source detection from the logarithm of the likelihood ratio when fitting the ROI with and without the putative dSph source (see Equation 6 in~\citealt{Ackermann:2015zua}).
We note that the \TS of an individual target does not depend on the assumed \Jfactor; however, both the \Jfactor and the \Jfactor uncertainty affect the DM interpretation.
No significant excess \gammaRayHyph emission above the background was observed coincident with any of the targets for any of the DM masses or channels tested.
Several of the targets show slight (${<}2.5\sigma$ local) excesses with respect to the background and are discussed further.

\begin{figure*}[t]
  \includegraphics[width=0.50\linewidth]{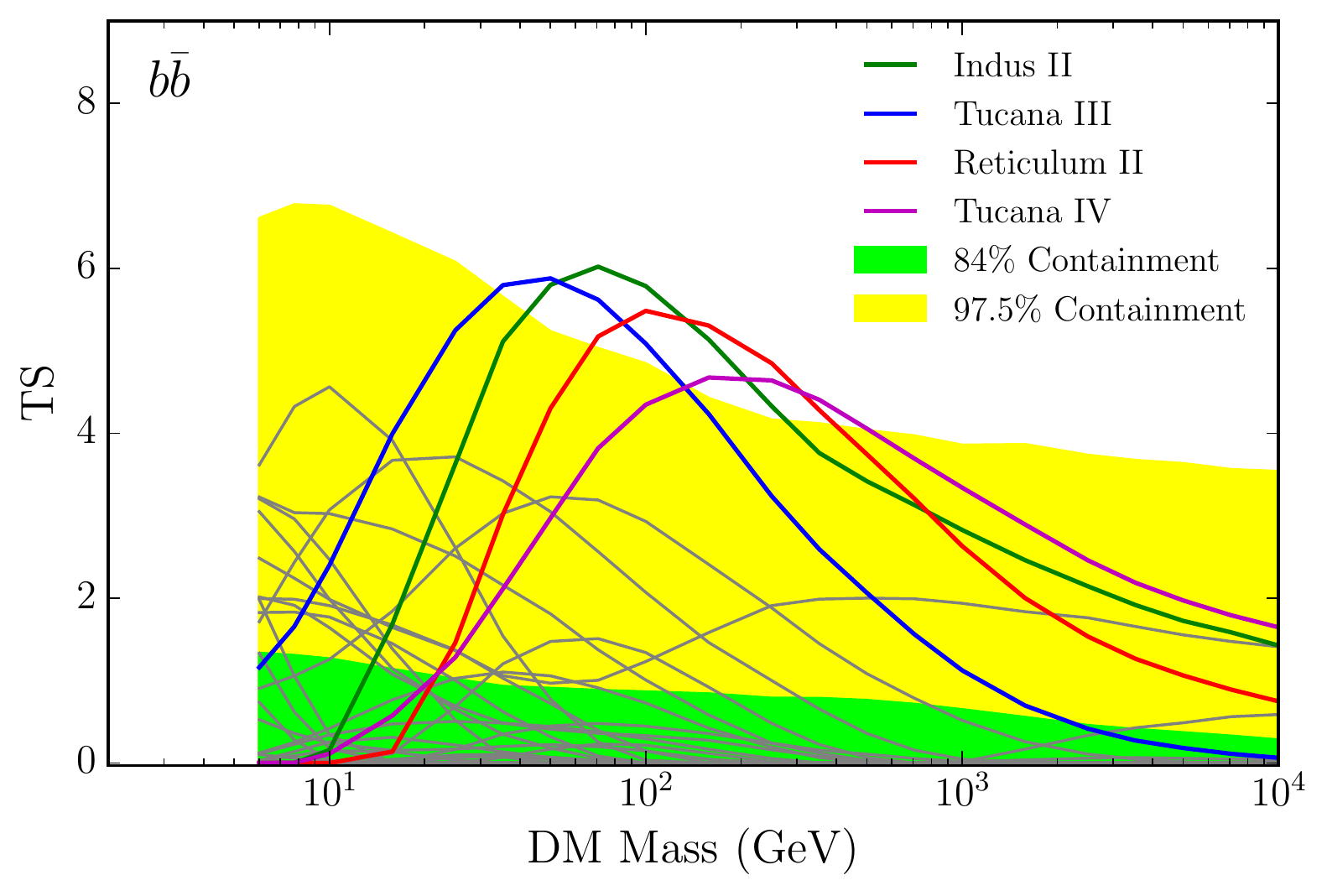}
  \includegraphics[width=0.50\linewidth]{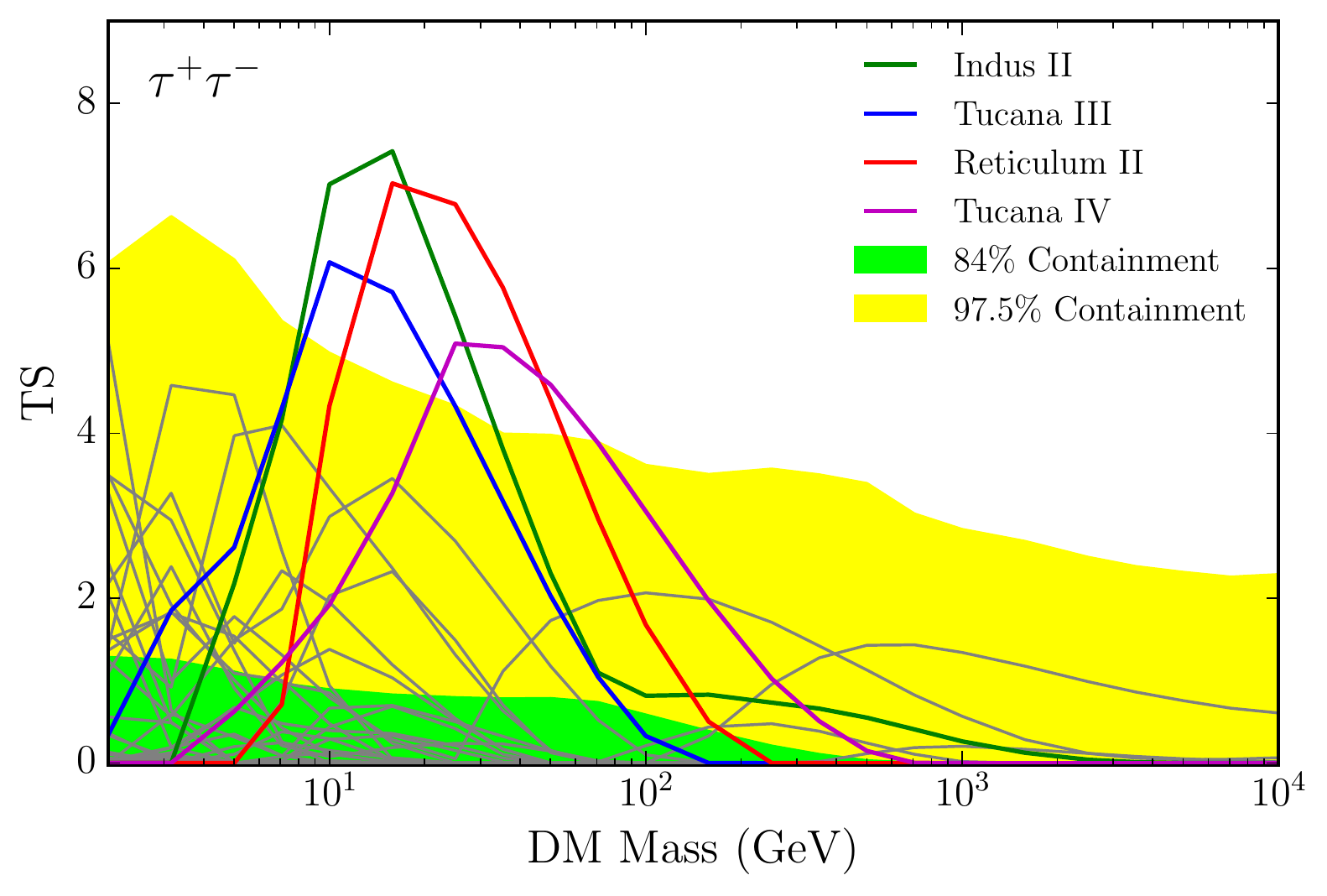}
  \caption{\label{fig:TSvsMass} Local detection significance,
    expressed as a log-likelihood test statistic (TS), from the
    broad-band analysis of each target in \tabref{targets} assuming DM
    annihilation through the \bbbar (left) or \tautau (right)
    channels.  The bands represent the local one-sided 84\% (green)
    and 97.5\% (yellow) containment regions derived from 300 random
    sets of 45 blank-sky locations.  Curves corresponding to targets
    with peak significance larger than the local 95\% expectation from
    blank-sky regions are explicitly colored and labeled, while other
    targets are shown in gray.  }
\end{figure*}

In \figref{TSvsMass} we show the \TS values from the likelihood analysis of each target as a function of annihilation channel and DM mass.
We also show the one-sided 84\% and 97.5\% containment bands from performing our analysis on blank-sky locations.\footnote{\NEW{The blank-sky locations used to calibrate detection significance are randomly distributed at Galactic latitudes $|b| > 30\degree$. The average diffuse background intensity in this region is within 5\% of the average diffuse background intensity in the 45 target ROIs. The incidence of unmodeled point sources is expected to be similar at blank-sky locations and in the target ROIs because the \gammaRayHyph sources detected at high Galactic latitudes are approximately isotropically distributed.}}
There are four targets with maximum \TS values exceeding the local 95\% containment contours from an analysis of blank-sky regions: \indII, \retII, \tucIII, and \tucIV.
\NEW{We note that other independent analyses have found significant (\plocal $>3\sigma$) emission from \retII~\citep{Geringer-Sameth:2015lua,Hooper:2015ula}.}
\NEW{The \plocal of \retII in this analysis is smaller mostly due to the use of the \irf{Pass 8} dataset as opposed to the \irf{Pass 7 Reprocessed} dataset.}

All four targets in \tabref{pvalue} have $\TS < 7.5$ when fit over the broad-band energy range with any DM spectral model ($\TS < 4$ when fit with a single $\Gamma = 2$ power-law spectral model).
The best-fit masses, channels, and significances of these excesses are shown in \tabref{pvalue}.
%We quote three \pvalues: (1) the local \pvalue, \plocal, calculated as the significance with respect 300 random sets of 45 blank-sky locations \citep{Ackermann:2013yva}, (2) the \pvalue per target, \pobject, which takes into account a trials factor from scanning multiple DM masses and channels \NEW{(also estimated empirically from random blank-sky locations)}, and (3) the sample \pvalue, \psample, which includes an additional trials factor from analyzing 45 target locations.
\NEW{ We quote three \pvalues: (1) the local \pvalue at the best-fit
  DM mass and channel, \plocal, (2) the \pvalue per target, \ptarget,
  which takes into account the trials factor from scanning multiple DM
  masses and channels, and (3) the sample \pvalue, \psample, which
  includes an additional trials factor from analyzing 45 target
  locations.  \plocal and \ptarget are empirically determined with
  respect to 300 sets of 45 blank-sky locations
  \citep{Ackermann:2013yva}.  For a particular target, the null
  distribution for \plocal is the distribution of TS evaluated at the
  best-fit DM mass and channel, whereas the null distribution for
  \ptarget is the distribution of the maximum TS over all considered
  DM masses and channels at each blank-sky location.
%, using the best-fit DM mass and channel for each target, or the maximum TS across DM masses and channels, respectively.
We use the TS distribution from fits in blank-sky locations to account for the effect of unmodeled components of the \gammaRayHyph sky such as unresolved point sources (see Figure 6 of the supplemental material for \citealt{Ackermann:2015zua}).
}

%We use the TS distribution from fits in blank-sky locations to calculate \pvalues to account for the effect of unmodeled components of the \gammaRayHyph sky such as unresolved point sources (see Figure 6 of the supplemental material for \citealt{Ackermann:2015zua}).

In the background-only case without a DM annihilation signal, analyzing 45 targets will yield four or more targets with detection significances exceeding the \pobject values in \tabref{pvalue} 45\% of the time.
However, this naive calculation treats each target equally, whereas the predicted \gammaRayHyph flux from DM annihilation is proportional to the \Jfactor.
In \secref{darkmatter}, we describe a combined analysis that weights the targets by their \Jfactors and links the spectral model (DM mass and annihilation channel) across targets, and thereby enhances the sensitivity to a collective DM signal from the population of Milky Way satellites.

No 3FGL sources are located within $1 \degr$ of any of the four systems mentioned above.
We also investigated associations with sources observed at other wavelengths that are potential \gammaRayHyph emitters in the BZCAT \citep{Massaro:2008ye}, CRATES \citep{Healey:2007by}, CGraBS \citep{Healey:2007gb}, PMN \citep{Wright:1994}, and WISE blazar candidate \citep{D'Abrusco:2014vba} catalogs.
We find two sources from the PMN catalog, PMN\,J0335$-$5406 and PMN\,J0335$-$5352, within $15 \arcmin$ of \retII.
The first of these, PMN\,J0335$-$5406, has a relatively large flux at low frequency (225\,mJy at 843\,MHz) and a fairly hard radio spectral index ($\Gamma \sim 0.7$), making it a possible \gammaRayHyph emitter \citep{Ackermann:2015yfk}.
In addition, the infrared colors of PMN\,J0335$-$5406 measured with WISE are consistent with other known \gammaRayHyph emitting blazars \citep{Massaro:2012dh}.
However, we note that this source is relatively faint in the optical/near infrared, having $z \gtrsim 23 \magn$ in the DES imaging.
The second source, PMN\,J0335$-$5352, has a smaller radio flux and seems unlikely to be associated with a \gammaRayHyph emitting blazar.
We additionally find the source PMN\,J0003$-$6059 located $10 \arcmin$ from \tucIV, but due to the lack of multifrequency measurements it is unclear whether it is a potential \gammaRayHyph emitter.

\renewcommand{\tblcaption}{\label{tab:pvalue} Targets with the Largest Excesses above Background}
\renewcommand{\tblcomments}{(1) Target name (2) best-fit DM annihilation channel (3) best-fit DM particle mass (4) highest TS value (5) local \pvalue calibrated from random blank regions (6) target \pvalue applying a trials factor from testing multiple DM annihilation spectra (7) sample \pvalue applying an additional trials factor from analyzing 45 targets. The Gaussian significance associated with each \pvalue is given in parentheses. More details can be found in \secref{LAT}.}
\begin{deluxetable*}{l ccc ccc}
\tablewidth{0.7\textwidth}
\tabletypesize{\small}
\tablecaption{ \tblcaption }
\tablehead{
(1) & (2) & (3) & (4) & (5) & (6) & (7) \\
Name & Channel & Mass (GeV) & \TS &  \plocal & \pobject & \psample }
\startdata
\indII                       & \tautau         & 15.8           & 7.4       & 0.01 (2.3$\sigma$) & 0.04 (1.7$\sigma$) & 0.84 (-1.0$\sigma$) \\
\retII                       & \tautau         & 15.8           & 7.0       & 0.01 (2.3$\sigma$) & 0.05 (1.7$\sigma$) & 0.88 (-1.2$\sigma$) \\
\tucIII                      & \tautau         & 10.0           & 6.1       & 0.02 (2.1$\sigma$) & 0.06 (1.5$\sigma$) & 0.94 (-1.6$\sigma$) \\
\tucIV                       & \tautau         & 25.0           & 5.1       & 0.02 (2.1$\sigma$) & 0.09 (1.3$\sigma$) & 0.98 (-2.1$\sigma$) 

\enddata
{\footnotesize \tablecomments{ \tblcomments }}
\end{deluxetable*}

\section{Estimating J-Factors}
\label{sec:jfactors}

An estimate of the \Jfactor is necessary to convert a \gammaRayHyph flux upper limit into a constraint on the DM annihilation cross section (Equation \ref{eqn:annihilation}).
The \Jfactor depends on both the DM density profile and distance.
Distances can be determined from the photometric data using the characteristic absolute magnitude of the main-sequence turn-off and/or horizontal branch in old, metal-poor stellar populations.
On the other hand, measurement of the DM mass requires spectroscopic observations to determine the radial velocities of member stars.
The classical dSphs discovered prior to SDSS have measured velocity dispersions in the range $\roughly 6 \text{--} 11 \kms$, and the ultra-faint dSphs discovered by SDSS have velocity dispersions in the range $\roughly 2 \text{--} 6 \kms$.
% Ret II = 3.3 km/s, Hor I = 4.9 km/s, Dra II = 2.9 km/s, Tri II = 4.9/5.1 km/s, Hya II = 5.4 km/s, Tuc II = 8.6 km/s
Similarly, the six new systems recently confirmed as dSphs have velocity dispersions in the range from $2.9 \text{--} 8.6 \kms$.

\begin{figure*}%[Th]
  \begin{centering}
\includegraphics[width=\textwidth]{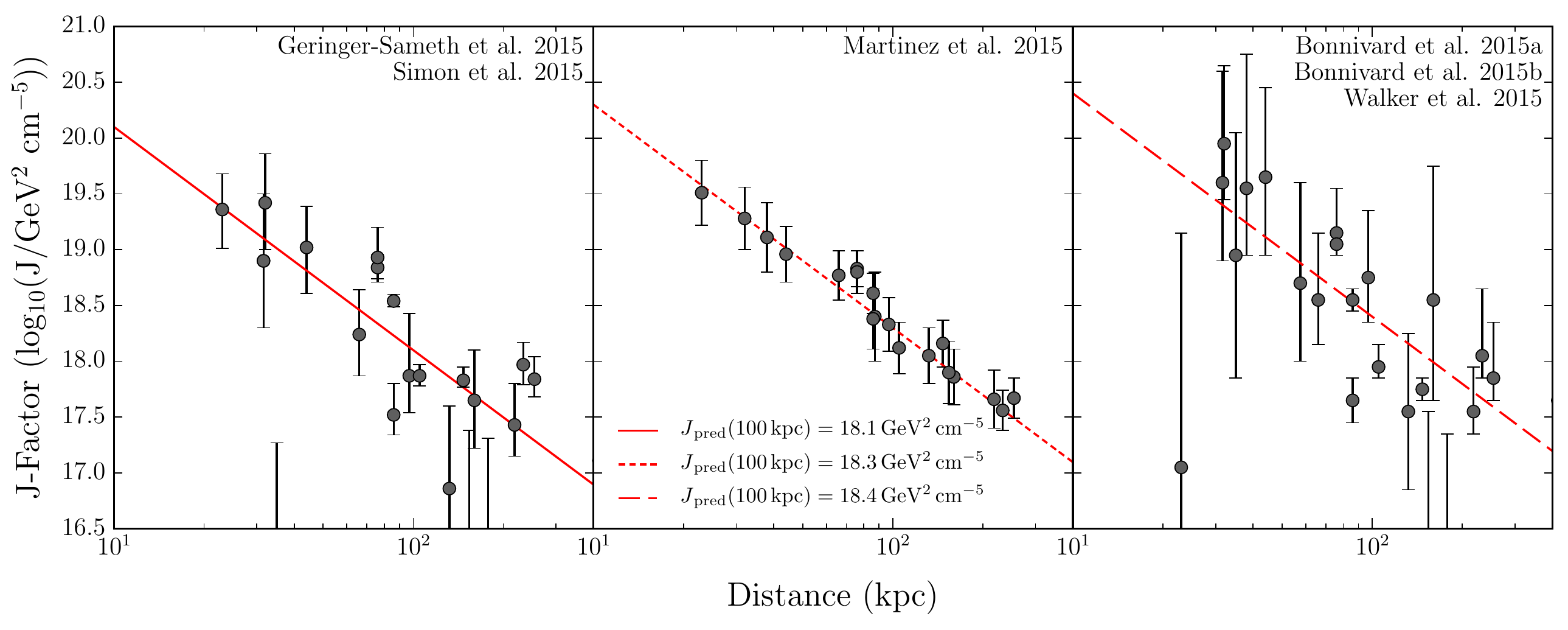}
  \caption{\label{fig:jfactors}
Relationship between the distances and spectroscopically determined \Jfactors of known dSphs is derived with three different techniques: ({\it left}) non-informative priors \citep{Geringer-Sameth:2014yza}, ({\it center}) Bayesian hierarchical modeling \citep{Martinez:2013els}, and ({\it right}) allowing for more flexible parametrizations of the stellar distribution and orbital anisotropy profile \citep{Bonnivard:2015xpq}. 
We also include recently derived \Jfactor estimates for \retII \citep{Simon:2015fdw,Bonnivard:2015tta} and \tucII \citep{2015arXiv151106296W} with \Jfactors for other dSphs that were calculated in a similar manner (see references for each panel).
We fit the \Jfactor scaling relation (Equation~\ref{eqn:jpred}) to the data in each panel, yielding $\log_{10}(J_0 / \GeV^{2} \cm^{-5}) = \{ 18.1, 18.3, 18.4 \}$, for the left, center, and right panels, respectively;
these relationships are plotted as solid, short dashed, and long dashed red lines.
}
\end{centering}
\end{figure*}

The known dSphs have similar central DM densities despite a wide spread in optical luminosity \citep{Strigari:2008ib}.
The similarity in the central DM density of the dSphs causes their \Jfactors to scale approximately as the inverse square of their distances.
In \figref{jfactors}, we show that a simple scaling relationship between \Jfactor and distance can be clearly seen in the \Jfactors derived by several groups \citep[\ie,][]{Geringer-Sameth:2014yza,Martinez:2013els,Bonnivard:2015xpq}.
\NEW{For each set of \Jfactor measurements, the intrinsic scatter relative to the proposed scaling relationship appears to be smaller than the average measurement uncertainty.}

Following DW15, we assume that the new stellar systems occupy similar DM halos to the population of known dSphs, and we predict the \Jfactors of the new systems from their distances.
This assumption is necessary to convert the \gammaRayHyph flux limits to DM annihilation cross section constraints since most of the newly discovered systems have not yet been observed spectroscopically.
We do not expect globular clusters to follow the same scaling relation, since their observed velocity dispersions imply that they do not contain DM.

For each candidate we calculated a predicted \Jfactor using the procedure developed in DW15.
Our scaling relationship is
\begin{equation}
\label{eqn:jpred}
%\log_{10}( J_{\rm pred} / J_0 ) = - 2 \log_{10} (D / 100 \kpc),
\log_{10} \left( \frac{J_{\rm pred}}{J_0} \right) = - 2 \log_{10} \left ( \frac{D}{100 \kpc} \right),
\end{equation}
\noindent where $D$ is the heliocentric distance of the dSph candidate and $J_0$ is a scale factor derived from a fit to spectroscopic data (\figref{jfactors}).
In contrast to DW15, we derived our nominal scale factor, $J_0 = 18.1 \GeV^2 \cm^{-5}$, using the spectroscopic \Jfactors from \citet{Geringer-Sameth:2014qqa} as opposed to those from \citet{Martinez:2013els}.
The two data sets give compatible results (see DW15); however, the \Jfactors derived by \citet{Geringer-Sameth:2014qqa} rely on fewer assumptions about the population of dSphs and provide slightly more conservative estimates for the predicted \Jfactors.
The predicted \Jfactor for each stellar system is shown in \tabref{targets}.

In addition to predicting the value of the \Jfactor we approximate the uncertainty achievable with future radial velocity measurements.
The uncertainty on the \Jfactor derived from spectroscopic observations depends on several factors, most importantly the number of stars for which radial velocities have been measured.
For ultra-faint dSphs that are similar to the dSph candidates, spectra have been measured for 20--100 stars.
Additional sources of uncertainty include the DM density profile and dynamical factors such as the velocity anisotropy of member stars.
We consider characteristic \Jfactor uncertainties, $\log_{10} \sigma_{J} = \{ 0.4, 0.6, 0.8\} \dex$, for the newly discovered ultra-faint satellites lacking spectroscopically determined \Jfactors.
Note that these uncertainties refer to characteristic measurement uncertainties on the \Jfactor for a typical dSph, and do not reflect any intrinsic scatter that may exist in a larger population of satellites.

We reiterate that this analysis assumes that the newly discovered systems are DM-dominated, similar to the known population of ultra-faint dSphs.
Some of the more compact systems might actually be faint outer-halo star clusters.
Some of the larger systems also may be subject to tidal stripping, in which case the distance-based estimation described above may not apply.
On-going spectroscopic analyses seek to robustly determine the DM content of new systems and identify those that have complicated kinematics.

\section{Dark Matter Constraints}
\label{sec:darkmatter}

\begin{figure*}[th]
  \begin{centering}
  \includegraphics[width=0.49\linewidth]{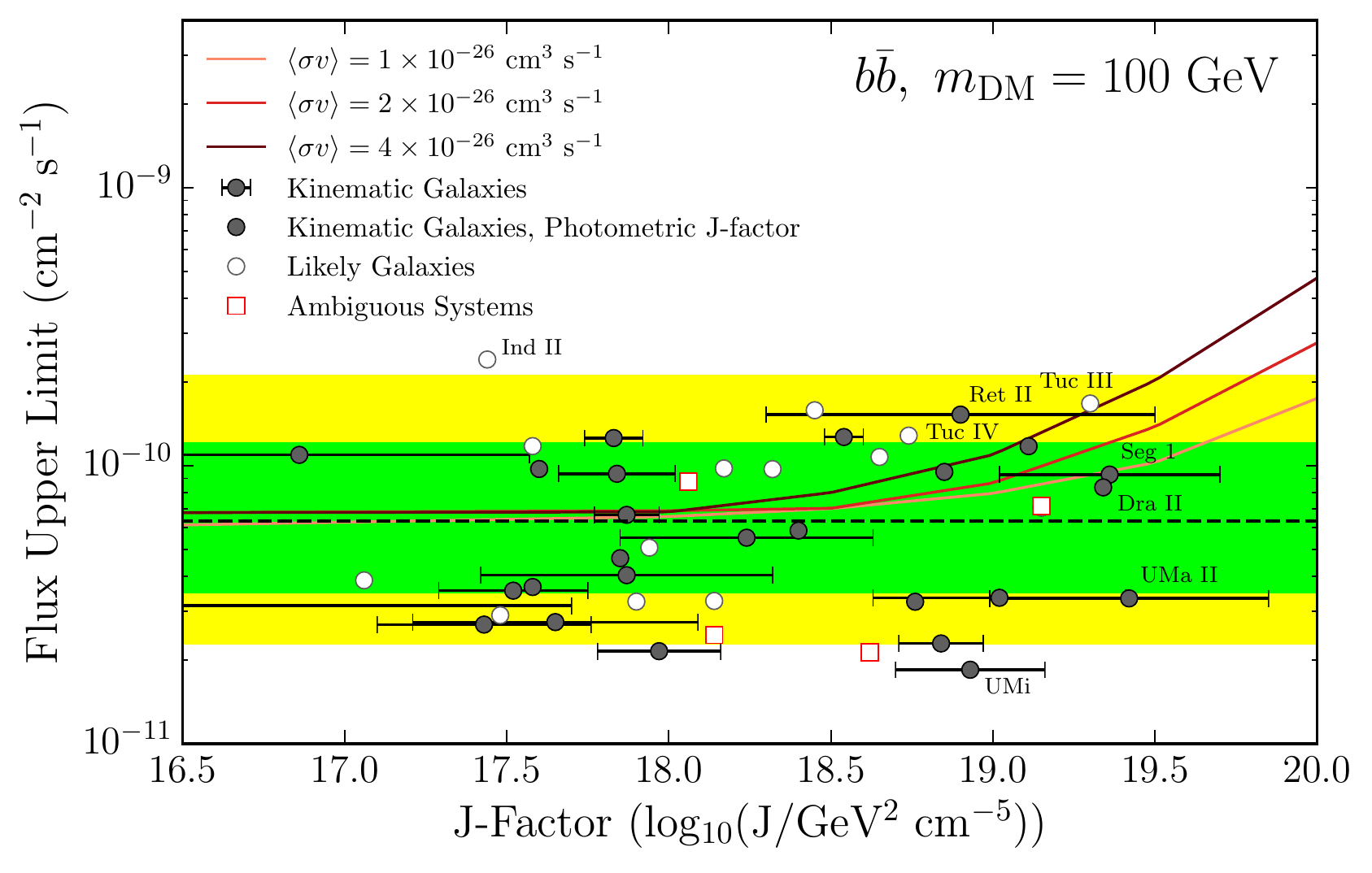}
  \includegraphics[width=0.49\linewidth]{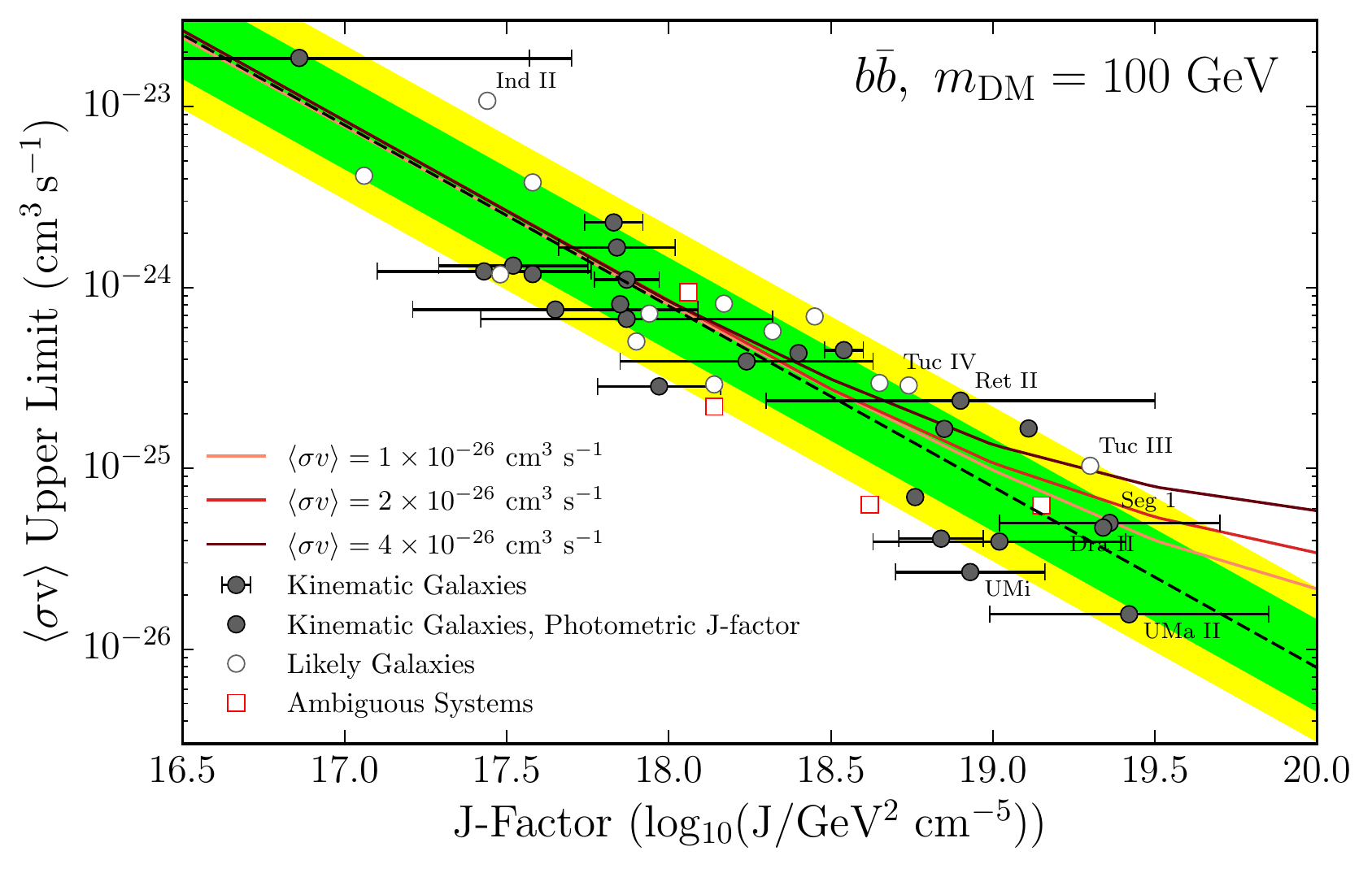}
  \caption{\label{fig:flux_vs_jfactor}
  Upper limits on flux ({\it left}) and cross section ({\it right}) versus \Jfactor.
  The points represent \Jfactors for each target estimated either from spectroscopy (filled circles with error bars) or from the scaling relation discussed in \secref{jfactors} (filled circles).
  The green and yellow shaded regions are the 68\% and 95\% containment regions for the blank-sky expectations, respectively.
  For comparison, the three solid lines show the median expected upper limits for DM annihilation with the given cross section.
  No significant deviation from the background-only expectation is observed.
    }
\end{centering}
\end{figure*}

We use the spectroscopically determined \Jfactors (when possible) and predicted \Jfactors (otherwise) for each confirmed and candidate dSph to interpret the \gammaRayHyph flux upper limits within a DM framework.
\figref{flux_vs_jfactor} summarizes the observed flux and \sigmav upper limits derived for individual confirmed and candidate dSphs, assuming a DM particle with a mass of $100 \GeV$ annihilating through the \bbbar-channel.\footnote{Results for both channels as well as bin-by-bin likelihood
  functions for each target are available in machine-readable format
  at: \url{http://www-glast.stanford.edu/pub_data/1203/}.}
We find that the observed upper limits are consistent with expectations from blank-sky regions.
We also show the median expected upper limit assuming that DM annihilates with a cross section comparable to the thermal relic cross section.
Targets with $\log_{10}(J/\GeV^2 \cm^{-5}) \lesssim 18.3$ would have a negligible \gammaRayHyph signal for a DM cross section similar to the thermal relic value.
However, the upper limits for systems with larger \Jfactors would be expected to deviate from the null hypothesis.

Given the large \Jfactors for \retII (measured) and \tucIII (predicted) we consider whether the low-significance excess emission observed toward them is consistent with a DM annihilation signal (\tabref{pvalue} and \figref{TSvsMass}).
Several other confirmed and candidate dSphs have \Jfactors comparable to \retII and \tucIII but have no excess over the background. 
The largest observed excess is associated with \indII, which, at a distance of $214 \kpc$, has a predicted signal that is two orders of magnitude smaller than for the most-promising candidates.
In addition, evidence for tidal tails associated with \tucIII \citep{Drlica-Wagner:2015b} might indicate that the DM halo of this stellar system is being tidally stripped.
Tidal stripping might significantly lower the \Jfactor of this target compared to the expectation when assuming hydrostatic equilibrium, which would decrease the predicted \gammaRayHyph flux from DM annihilation.
Finally, based on an analysis of blank-sky regions and the number of targets considered, a maximum TS value at least as large as that observed is expected in the majority of background-only realizations (see column 7 of \tabref{pvalue}).

To further explore the consistency of the \gammaRayHyph data with a DM annihilation signal from the dSph population, and to increase search sensitivity, we combined observations of multiple satellite systems in a joint likelihood analysis.
By simultaneously analyzing the population of confirmed and candidate dSphs, we avoid a look-elsewhere effect from focusing on excesses or deficits associated  with individual targets.
As opposed to weighting each target equally, the combined likelihood analysis emphasizes those targets with the largest \Jfactors and enforces consistency in the DM annihilation spectrum.

The current uncertainty in the photometric classification of newly found systems motivates the definition of three target samples for our combined analysis (\tabref{targets}).
\begin{enumerate}
\item Our ``nominal'' sample includes: (1) kinematically confirmed dSphs, and (2) systems with  $r_{1/2} > 20 \pc$ and $\mu > 25 \magn \unit{arcsec}^{-2}$.
\item We defined a ``conservative'' sample as a sub-selection of the nominal sample excluding systems with kinematic or photometric indications of tidal disruption.
Specifically, the conservative sample excludes \booIII and \wilI, which appear to be dSphs but have kinematics that are difficult to interpret \citep{Carlin:2009cu,Willman:2010gy}.
Additionally, we exclude the new system \tucIII, which shows possible indication of tidal stripping \citep{Drlica-Wagner:2015b}.
\item Finally, we define an ``inclusive'' sample, which augments the nominal sample selection with all systems with  $r_{1/2} > 10 \pc$ and $\mu > 25 \magn \unit{arcsec}^{-2}$.
This sample includes four ambiguous systems: \cetII, \eriIII, \kimII, and \tucV.
\end{enumerate}
These sample selections are compared to the photometric characteristics of dSphs and globular clusters in \figref{size_luminosity} and are indicated in \tabref{targets}.

\begin{figure*}[t]
  \includegraphics[width=0.5\linewidth]{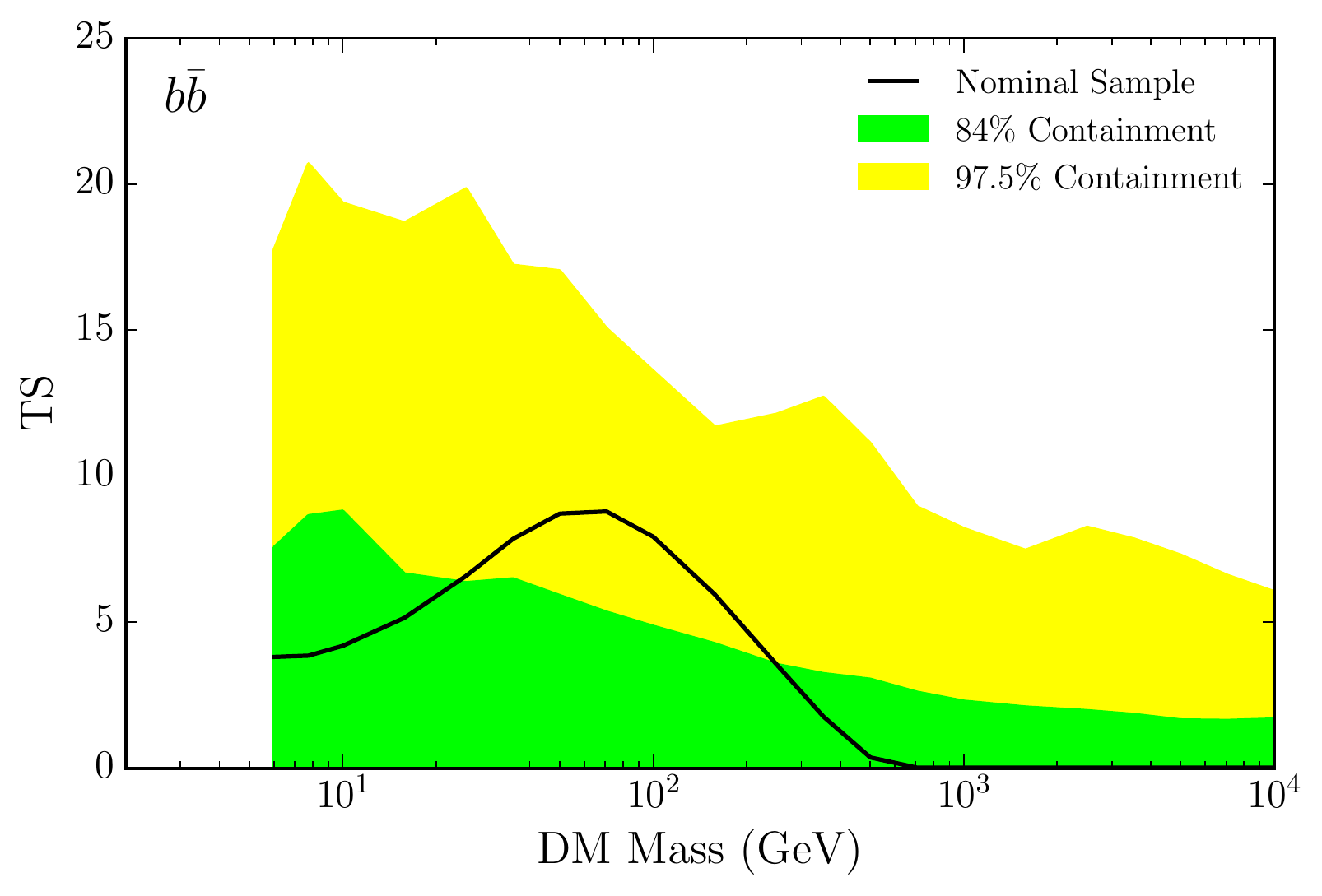}
  \includegraphics[width=0.5\linewidth]{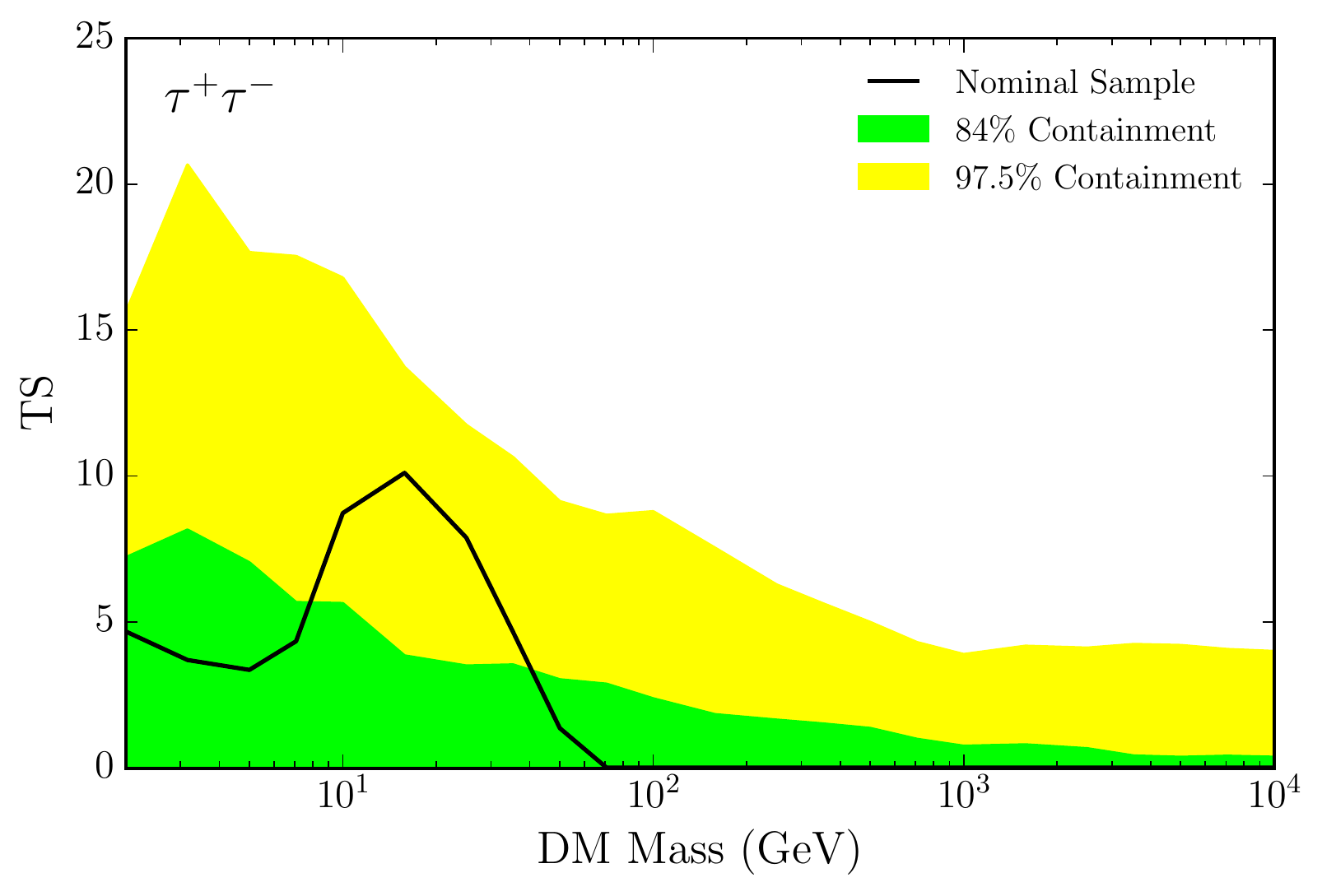}
  \caption{\label{fig:resultsTS} \NEW{Local detection significance,
    expressed as a log-likelihood test statistic (TS), from the
    combined analysis of the nominal target sample assuming DM
    annihilation through the \bbbar ({\it left}) or \tautau ({\it
      right}) channels. The log-normal \Jfactor uncertainties for targets lacking 
    spectroscopic \Jfactors are 0.6 dex in this example. The bands represent the local one-sided 84\%
    (green) and 97.5\% (yellow) containment regions derived from 300
    random sets of 45 blank-sky locations using the same set of \Jfactors 
    as in the nominal sample.}}
\end{figure*}

When analyzing the \gammaRayHyph data in the context of DM
annihilation, we made use of measured \Jfactors based on spectroscopic
observations when possible.  If spectroscopic \Jfactors were
unavailable, we used the values predicted from the distance scaling
relationship and adopted a nominal uncertainty of 0.6 dex.  We
followed the prescription of \citet{Ackermann:2015zua} to incorporate
the \Jfactor uncertainty as a nuisance parameter (see Equations 3--5
in \citealt{Ackermann:2015zua}).  The largest excess found in the
combined analysis of our nominal sample was \NEW{$\TS = 10.1$} for a
DM particle mass of 15.8 GeV annihilating into $\tau$-leptons (see
\NEW{\figref{resultsTS}}).  We calibrated this \TS against a sample of
randomly selected blank-sky locations to get
\NEW{$\plocal = 0.047\ (1.7 \sigma)$}.  We converted this to
\NEW{$\pglobal = 0.23\ (0.7\sigma)$} by applying a trials factor to
account for our scan in DM mass and annihilation channel.%
\footnote{If we only tested the single DM model best-fit to the GCE
  then it would not be necessary to include a trials factor for
  testing multiple DM masses and
  channels~\citep[\eg,][]{Hooper:2015ula}.}

%%%%%%%%%%%%%%%%%%%%%%%%%%%%%%%%%%%%%%%%
%%% ADD EXTENDED SOURCE RESULTS HERE %%%
%%%%%%%%%%%%%%%%%%%%%%%%%%%%%%%%%%%%%%%%
\citet{Ackermann:2013yva} found that cross section upper limits
derived from dSphs are fairly insensitive to the assumed spatial
extension.  However, we investigate the impact of modeling the targets
as spatially extended sources using the Navarro-Frenk-White (NFW) DM
density profiles projected along the line of sight
\citep{Navarro:1996gj}.  Since the scale radii of the dSph candidates
are not well constrained, we consider characteristic scale radii of
100\pc, 316\pc, and 1\kpc.  When assuming the largest scale radius of
1\kpc, we find that the TS of the most significant excess observed in
the analysis of the nominal sample (\tautau channel and mass of
15.8\GeV) increases to \NEW{$\TS = 15.3$}.  The global significance of
the excess assuming the most extended spatial model is
\NEW{$\pglobal = 0.21\ (0.8 \sigma)$}; however, this value does not
account for the additional trials factor from testing multiple spatial
models.

% Updated discussion moving away from the emphasis on TS values since
% we now better appreciate that the null distribution of TS depends on 
We also performed our analysis using predicted \Jfactor uncertainties
of 0.4 dex and 0.8 dex when spectroscopic \Jfactors were unavailable.
\NEW{The \TS values and associated detection significances from these analyses are listed in \tabref{results}.
Different choices for the target sample and predicted \Jfactor uncertainties yield distinct null distributions for the \TS.
The resulting \pglobal values do not account for the extra trials factor from testing multiple target
samples and \Jfactor uncertainties.
In all cases, $\pglobal < 1 \sigma$.}
Due to the lack of a significant excess in the combined analysis, we
conclude that there is no significant evidence of DM annihilation in
the population of confirmed and candidate dSphs.

%We also performed our analysis using predicted \Jfactor uncertainties
%of 0.4 dex and 0.8 dex when spectroscopic \Jfactors were unavailable.
%We present the \TS values from these analyses in \tabref{results}.
%The largest \TS (\NEW{$\TS = 11.6$}) is found for the inclusive sample
%assuming a \Jfactor uncertainty of \NEW{0.8} dex, which corresponds to
%\NEW{$\pglobal = 0.34\ (0.4 \sigma)$}.  This global \pvalue does not
%account for the extra trials factor from testing multiple target
%samples and \Jfactor uncertainties.  \NEW{Smaller} \Jfactor
%uncertainties and a conservative sample selection result in a
%decreased composite \NEW{TS}.
%%and a shift to a lower best-fit DM mass.  This behavior
%%is consistent with the results of \citet{Ackermann:2015zua}, who found
%%a maximum excess with $\TS = 1.3$ associated with $\mDM = 2 \GeV$
%%annihilating to \ee.  
%Due to the lack of a significant excess in the combined analysis, we
%conclude that there is no significant evidence of DM annihilation in
%the population of confirmed and candidate dSphs.

\renewcommand{\tblcaption}{\label{tab:results}
Combined Analysis Results
}
\renewcommand{\tblcomments}{
Largest \TS values from the combined analysis of satellite systems in our three target samples.
We adopt log-normal \Jfactor uncertainties of 0.4 dex, 0.6 dex, and 0.8 dex for targets lacking spectroscopic \Jfactors.
The global \pvalue is calibrated from random blank-sky regions and is corrected for a trials factor from fitting multiple DM annihilation spectra.
}
\begin{deluxetable}{llccl}
\tablewidth{0.45\textwidth}
\tabletypesize{\small}
\tablecaption{ \tblcaption }
\tablehead{
Sample & Channel  & Mass (GeV) & TS & \pglobal
}
\startdata

\multicolumn{5}{c}{ \ruleline{ 0.4 dex } }\\
Inclusive                    & \tautau         & 15.8           & 8.5        & 0.20 (0.8$\sigma$) \\
Nominal                      & \tautau         & 15.8           & 8.5        & 0.18 (0.9$\sigma$) \\
Conservative                 & \tautau         & 15.8           & 2.5        & 0.51 (-0.0$\sigma$) \\
\multicolumn{5}{c}{ \ruleline{ 0.6 dex } }\\
Inclusive                    & \tautau         & 15.8           & 10.1       & 0.27 (0.6$\sigma$) \\
Nominal                      & \tautau         & 15.8           & 10.1       & 0.23 (0.7$\sigma$) \\
Conservative                 & \tautau         & 15.8           & 3.0        & 0.60 (-0.3$\sigma$) \\
\multicolumn{5}{c}{ \ruleline{ 0.8 dex } }\\
Inclusive                    & \tautau         & 15.8           & 11.6       & 0.34 (0.4$\sigma$) \\
Nominal                      & \tautau         & 15.8           & 11.4       & 0.29 (0.6$\sigma$) \\
Conservative                 & \tautau         & 25.0           & 3.8        & 0.68 (-0.5$\sigma$)

\enddata
{\footnotesize \tablecomments{ \tblcomments }}
\end{deluxetable}

\begin{figure*}[t]
  \includegraphics[width=0.5\linewidth]{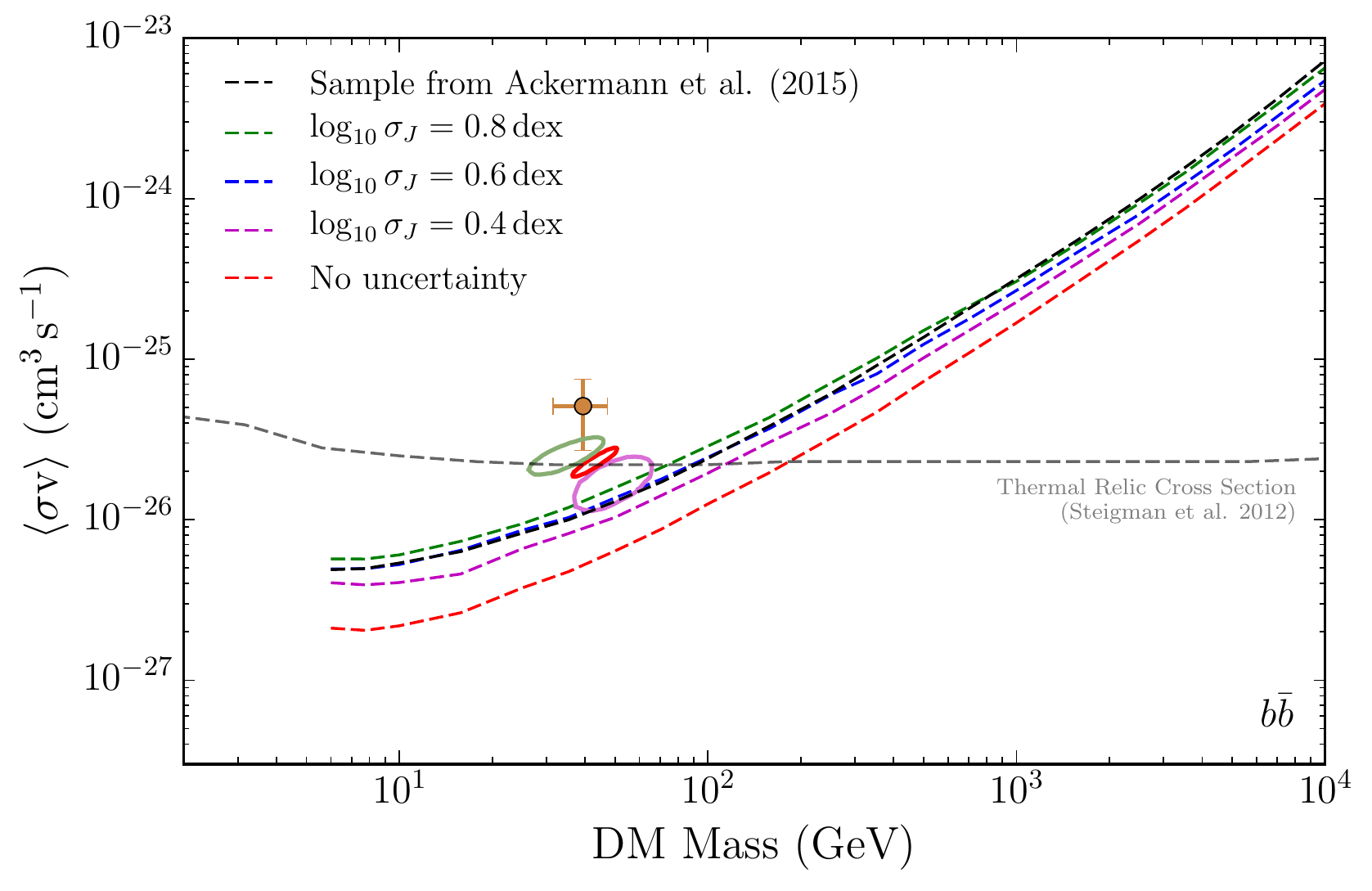} 
  \includegraphics[width=0.5\linewidth]{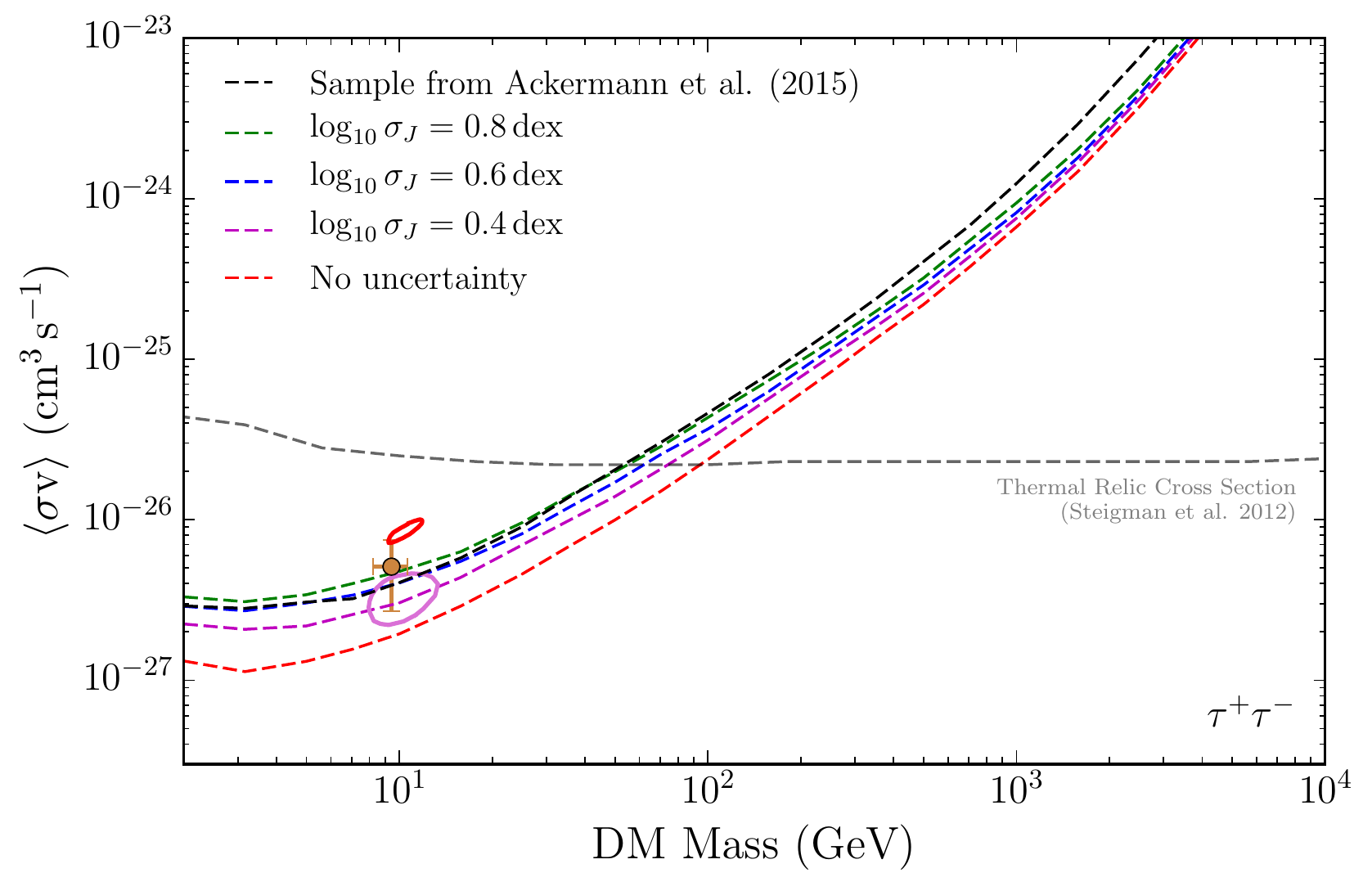} 
  \caption{\label{fig:expectUL}
  Expected sensitivity expressed as a limit on the DM annihilation
  cross section for the \bbbar ({\it left}) and \tautau ({\it right})
  channels.  The expected sensitivity is calculated as the median 95\%
  confidence level upper limit from 300 sets of random blank-sky
  locations.  The dashed black line shows the median expected
  sensitivity for the sample of 15 dSphs with kinematic \Jfactors used
  in the combined analysis of \citet{Ackermann:2015zua}.  Colored dashed
  curves show the median sensitivity for the combined analysis
  of the nominal sample derived assuming \Jfactor uncertainties of 0.8
  dex, 0.6 dex, and 0.4 dex for the targets with
  distance-based \Jfactor estimates.  The ``No Uncertainty''
  expectation curve is derived assuming zero \Jfactor uncertainty for
  all targets and represents the limiting sensitivity attainable by
  reducing \Jfactor uncertainties.  The closed contours and marker show
  the best-fit regions (at $2\sigma$ confidence) in cross-section and
  mass from several DM interpretations of the
  GCE: green contour~\citep{Gordon:2013vta}, red contour~\citep{Daylan:2014rsa}, orange data point~\citep{Abazajian:2014fta}, purple contour~\citep{Calore:2014xka}. 
The dashed gray curve corresponds to the thermal relic cross section from \citet{Steigman:2012nb}.
}
\end{figure*}

\NEW{Assuming that the \Jfactors are an accurate representation of the expected dark matter annihilation signal, a combined analysis of the satellite population is more sensitive than the analysis of any individual target.}
In \figref{expectUL}, we show the median expected sensitivity for an analysis of our nominal sample assuming several different \Jfactor uncertainties for targets without spectroscopically determined \Jfactors (kinematic \Jfactors are held fixed in each case).
Additionally, we show the optimistic scenario where the \Jfactors for the entire sample can be determined exactly.
In this limiting case, the analysis is sensitive to the thermal relic cross section for DM particles with mass $\lesssim 200 \GeV$, a factor of $\roughly 2$ increase in mass relative to the analysis of \citet{Ackermann:2015zua}.

\begin{figure*}[t]
  \includegraphics[width=0.5\linewidth]{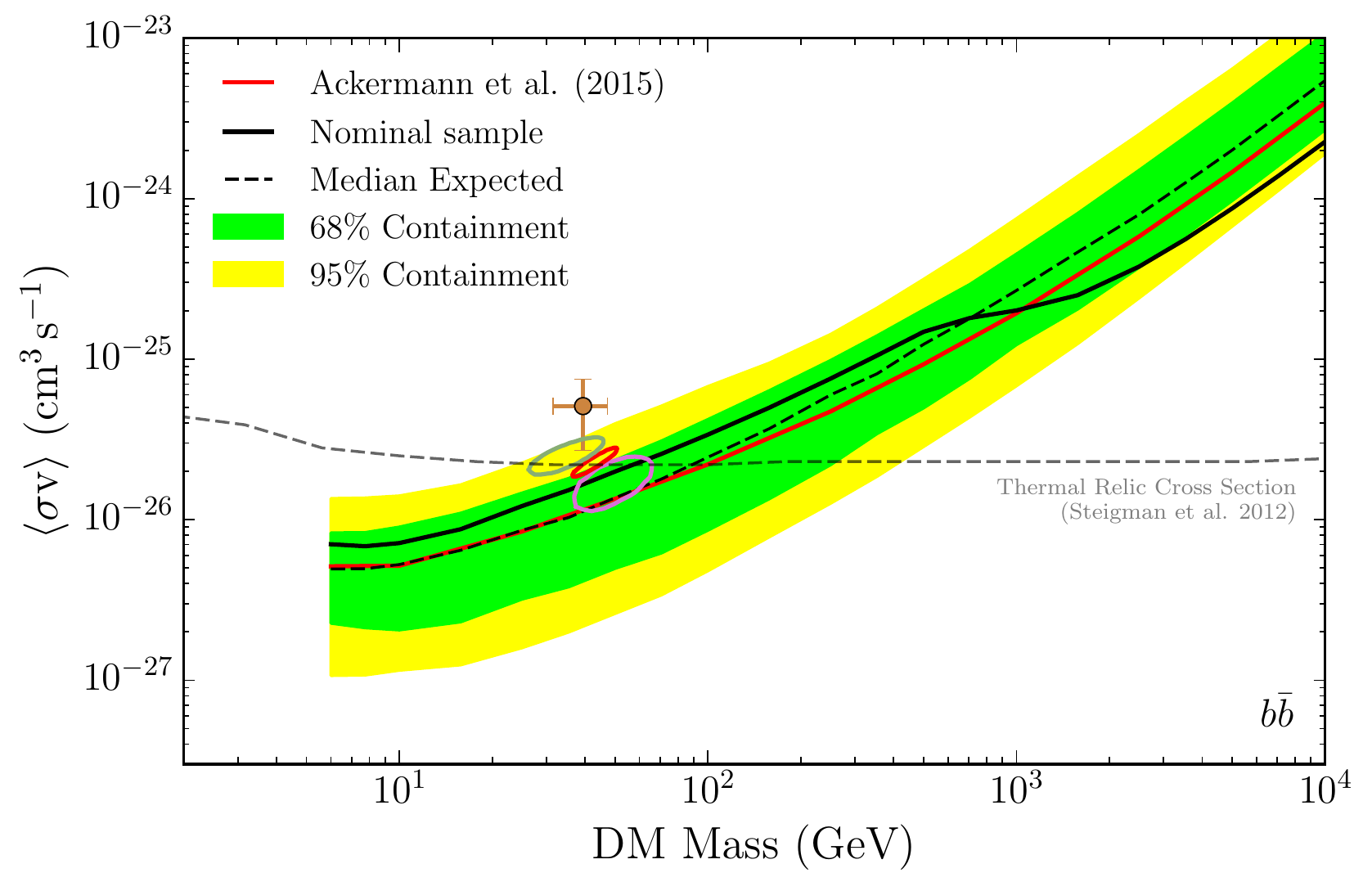}
  \includegraphics[width=0.5\linewidth]{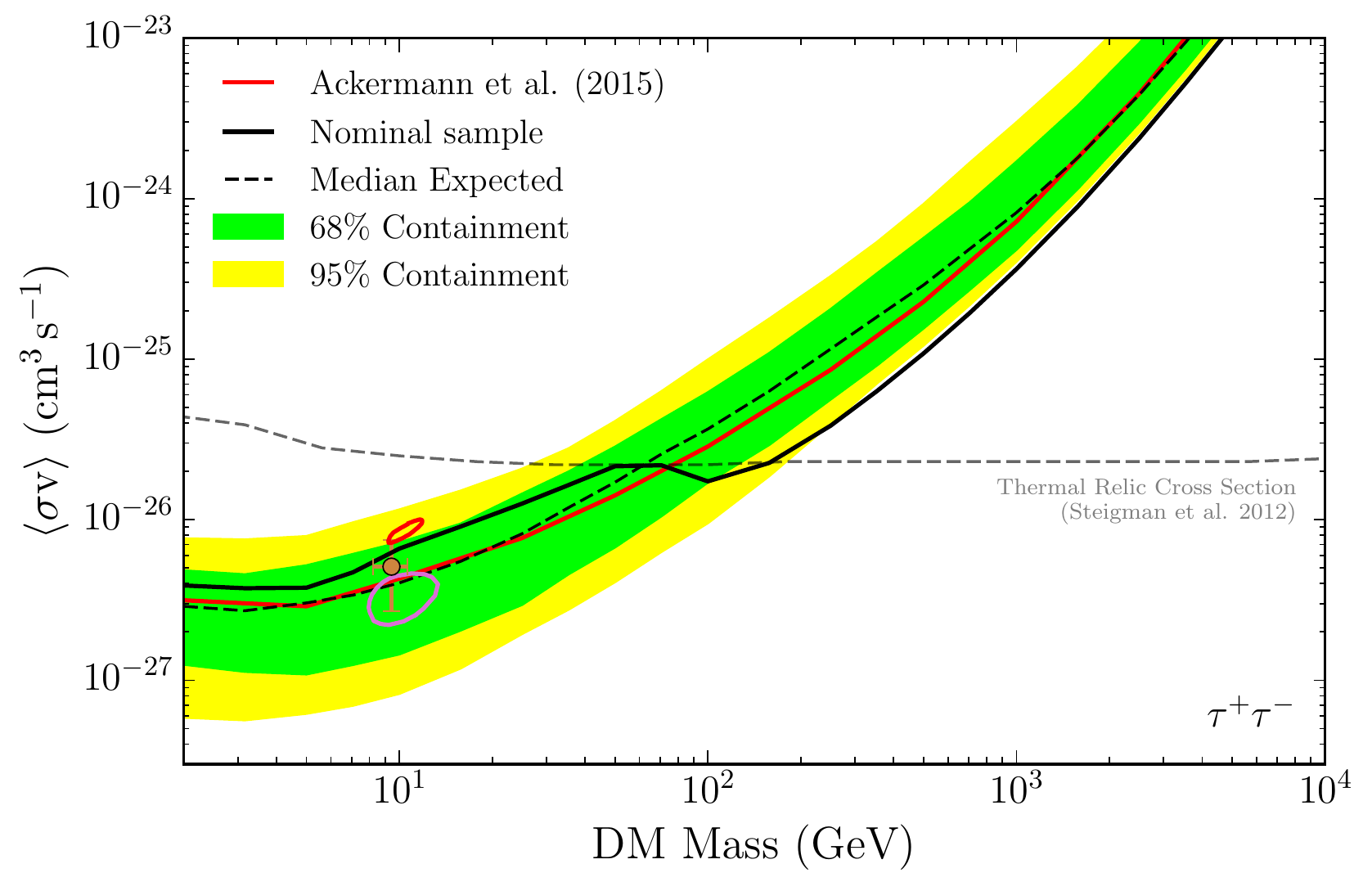}
  \caption{\label{fig:resultsUL} Upper limits (95\% confidence level)
    on the DM annihilation cross section derived from a combined
    analysis of the nominal target sample for the \bbbar ({\it left})
    and \tautau ({\it right}) channels.  Bands for the expected
    sensitivity are calculated by repeating the same analysis on 300
    randomly selected sets of high-Galactic-latitude blank fields in
    the LAT data.  The dashed line shows the median expected
    sensitivity while the bands represent the 68\% and 95\% quantiles.
    Spectroscopically measured \Jfactors are used when available;
    otherwise, \Jfactors are predicted photometrically with an
    uncertainty of 0.6 dex (solid red line).  The solid black line
    shows the observed limit from the combined analysis of 15 dSphs
    from \citet{Ackermann:2015zua}. 
    %The closed contours and marker are the same as depicted in \figref{expectUL}.
    The closed contours and marker
    show the best-fit regions (at $2\sigma$ confidence) in
    cross-section and mass from several DM interpretations of the GCE:
    green contour~\citep{Gordon:2013vta}, red
    contour~\citep{Daylan:2014rsa}, orange data
    point~\citep{Abazajian:2014fta}, purple
    contour~\citep{Calore:2014xka}.  The dashed gray curve corresponds
    to the thermal relic cross section from \citet{Steigman:2012nb}.
  }
\end{figure*}

In \figref{resultsUL} we show upper limits derived from a combined analysis of our nominal sample assuming a \Jfactor uncertainty of 0.6 dex for targets lacking spectroscopic \Jfactors.
We find that the derived upper limits are consistent within the range of statistical fluctuation expected from 300 random high-latitude blank-sky fields.
The derived upper limits lie above the median expectation for masses below $\roughly 1 \TeV$ and $\roughly 70 \GeV$ for the \bbbar and \tautau channels, respectively.
This behavior can be attributed to the low-significance excesses discussed in \secref{LAT}.
In contrast, we note that the limits lie below the median expectation at higher masses.
%This behavior might result from the fact that most of the Milky Way satellites reside outside the \Fermi Bubbles \citep{Su:2010qj,Ackermann:2014sfa} and are subject to a slightly lower high-energy diffuse background flux than the average high-latitude field.

\section{Conclusions}
\label{sec:conclusions}

We have performed a \NEW{comprehensive} \gammaRayHyph analysis of
\Fermi-LAT data coincident with 45 confirmed and candidate dSphs.  We
find no statistically significant \NEW{($>3\sigma$)} \gammaRayHyph excesses
toward any of our targets.  Four of the targets (including two nearby
systems) exhibit small excesses with local significances
$<2.5\sigma$.  
%The most significant target is at a distance of $>200 \kpc$ and is not expected to have a large DM annihilation signal.  
Since the characteristics of the DM particle (\ie, mass and
annihilation channel) are expected to be the same in all dSphs, we
perform a combined analysis on the sample of confirmed and candidate
dSphs.  We use a simple scaling relationship to predict the DM
annihilation signal in systems without spectroscopic data.  When
considering the ensemble of targets, the \gammaRayHyph data are
consistent with the background-only null hypothesis.  The maximum
excess found in a joint likelihood analysis of our nominal target
sample yields a maximum global significance of
\NEW{$\pglobal = 0.23\ (0.7\sigma)$} for a DM mass of
\CHECK{$15.8 \GeV$} annihilating via the \tautau channel.

We calculate the median expected sensitivity assuming the DM contents of the new candidate dSphs are comparable to those of previously known dSphs.
The expected sensitivity to DM annihilation improves as more targets are added, and depends on the precision with which the \Jfactors of the new systems can be measured, as well as the DM mass and annihilation channel being tested.
Assuming that the \Jfactors of the new systems can be measured with an uncertainty of 0.6 dex, the improvement in sensitivity is a factor of $\roughly 1.5$ for hard annihilation spectra (\eg, the \tautau channel) compared to the median expected limits in \citet{Ackermann:2015zua}.
More precisely determined \Jfactors are expected to improve the sensitivity by up to a factor of 2, motivating deeper spectroscopic observations both with current facilities and future thirty-meter class telescopes \citep{GMT:2014,TMT:2015}.

The limits derived from LAT data coincident with confirmed and candidate dSphs do not yet conclusively confirm or refute a DM interpretation of the GCE \citep{Gordon:2013vta,Daylan:2014rsa,Abazajian:2014fta,Calore:2014xka}.
Relative to the combined analysis of \citet{Ackermann:2015zua}, the limits derived here are up to a factor of 2 more constraining at large DM masses ($m_{{\rm DM}, \bbbar} \gtrsim 1\TeV$ and $m_{{\rm DM}, \tautau} \gtrsim 70 \GeV$) and a factor of $\roughly 1.5$ less constraining for lower DM masses.
The weaker limits obtained at low DM mass can be attributed to low-significance excesses coincident with some of the nearby and recently discovered stellar systems, \ie, \retII and \tucIII.
While the excesses associated with these targets are broadly consistent with the DM spectrum and cross section fit to the GCE, we refrain from a more extensive DM interpretation due to the low significance of these excesses, the uncertainties in the \Jfactors of these targets, and the lack of any significant signal in the combined analysis.

Ongoing \Fermi-LAT observations, more precise \Jfactor determinations with deeper spectroscopy, and searches for new dSphs in large optical surveys will each contribute to the future sensitivity of DM searches using Milky Way satellites \citep{Charles:2016pgz}.
In particular, the Large Synoptic Survey Telescope \citep{Ivezic:2008fe} is expected to find hundreds of new Milky Way satellite galaxies \citep{Tollerud:2008ze,Hargis:2014kaa}.
Due to the difficulty in acquiring spectroscopic observations and the relative accessibility of \gammaRayHyph observations, it seems likely that \gammaRayHyph analysis will precede \Jfactor determinations in many cases.
To facilitate updates to the DM search as spectroscopic \Jfactors become available, the likelihood profiles for each energy bin used to derive our \gammaRayHyph flux upper limits will be made publicly available.
We plan to augment this resource as more new systems are discovered.

After the completion of this analysis, we became aware of an independent study of LAT \irf{Pass 8} data coincident with DES Y2 dSph candidates \citep{Li:2015kag}.
The \gammaRayHyph results associated with individual targets are consistent between the two works; however, the samples selected for combined analysis are different.

\section*{Acknowledgments}

\NEW{We would like to thank Tim Linden and Dan Hooper for helpful and engaging conversations.
We also thank the anonymous referee for thoughtful and constructive feedback on this manuscript.}

The \textit{Fermi} LAT Collaboration acknowledges generous ongoing support from a number of agencies and institutes that have supported both the development and the operation of the LAT as well as scientific data analysis.
These include the National Aeronautics and Space Administration and the Department of Energy in the United States, the Commissariat \`a l'Energie Atomique and the Centre National de la Recherche Scientifique / Institut National de Physique Nucl\'eaire et de Physique des Particules in France, the Agenzia Spaziale Italiana and the Istituto Nazionale di Fisica Nucleare in Italy, the Ministry of Education, Culture, Sports, Science and Technology (MEXT), High Energy Accelerator Research Organization (KEK) and Japan Aerospace Exploration Agency (JAXA) in Japan, and the K.~A.~Wallenberg Foundation, the Swedish Research Council and the Swedish National Space Board in Sweden.
Additional support for science analysis during the operations phase is gratefully acknowledged from the Istituto Nazionale di Astrofisica in Italy and the Centre National d'\'Etudes Spatiales in France.

Funding for the DES Projects has been provided by the U.S. Department of Energy, the U.S. National Science Foundation, the Ministry of Science and Education of Spain,
the Science and Technology Facilities Council of the United Kingdom, the Higher Education Funding Council for England, the National Center for Supercomputing
Applications at the University of Illinois at Urbana-Champaign, the Kavli Institute of Cosmological Physics at the University of Chicago,
the Center for Cosmology and Astro-Particle Physics at the Ohio State University,
the Mitchell Institute for Fundamental Physics and Astronomy at Texas A\&M University, Financiadora de Estudos e Projetos,
Funda{\c c}{\~a}o Carlos Chagas Filho de Amparo {\`a} Pesquisa do Estado do Rio de Janeiro, Conselho Nacional de Desenvolvimento Cient{\'i}fico e Tecnol{\'o}gico and
the Minist{\'e}rio da Ci{\^e}ncia, Tecnologia e Inova{\c c}{\~a}o, the Deutsche Forschungsgemeinschaft and the Collaborating Institutions in the Dark Energy Survey.
The DES data management system is supported by the National Science Foundation under Grant Number AST-1138766.
The DES participants from Spanish institutions are partially supported by MINECO under grants AYA2012-39559, ESP2013-48274, FPA2013-47986, and Centro de Excelencia Severo Ochoa SEV-2012-0234,
some of which include ERDF funds from the European Union.

The Collaborating Institutions are Argonne National Laboratory, the University of California at Santa Cruz, the University of Cambridge, Centro de Investigaciones En{\'e}rgeticas,
Medioambientales y Tecnol{\'o}gicas-Madrid, the University of Chicago, University College London, the DES-Brazil Consortium, the University of Edinburgh,
the Eidgen{\"o}ssische Technische Hochschule (ETH) Z{\"u}rich,
Fermi National Accelerator Laboratory, the University of Illinois at Urbana-Champaign, the Institut de Ci{\`e}ncies de l'Espai (IEEC/CSIC),
the Institut de F{\'i}sica d'Altes Energies, Lawrence Berkeley National Laboratory, the Ludwig-Maximilians Universit{\"a}t M{\"u}nchen and the associated Excellence Cluster Universe,
the University of Michigan, the National Optical Astronomy Observatory, the University of Nottingham, The Ohio State University, the University of Pennsylvania, the University of Portsmouth,
SLAC National Accelerator Laboratory, Stanford University, the University of Sussex, and Texas A\&M University.

This research has made use of the NASA/IPAC Extragalactic Database (NED) which is operated by the Jet Propulsion Laboratory, California Institute of Technology, under contract with the National Aeronautics and Space Administration.

{\it Facilities:} \facility{Fermi}, \facility{Blanco}

\bibliographystyle{apj}
\bibliography{bib}

\cleardoublepage
\appendix

\subsection{Influence of Predicted J-factors}

\begin{figure*}[h]
  \centering 
  \includegraphics[width=0.49\textwidth]{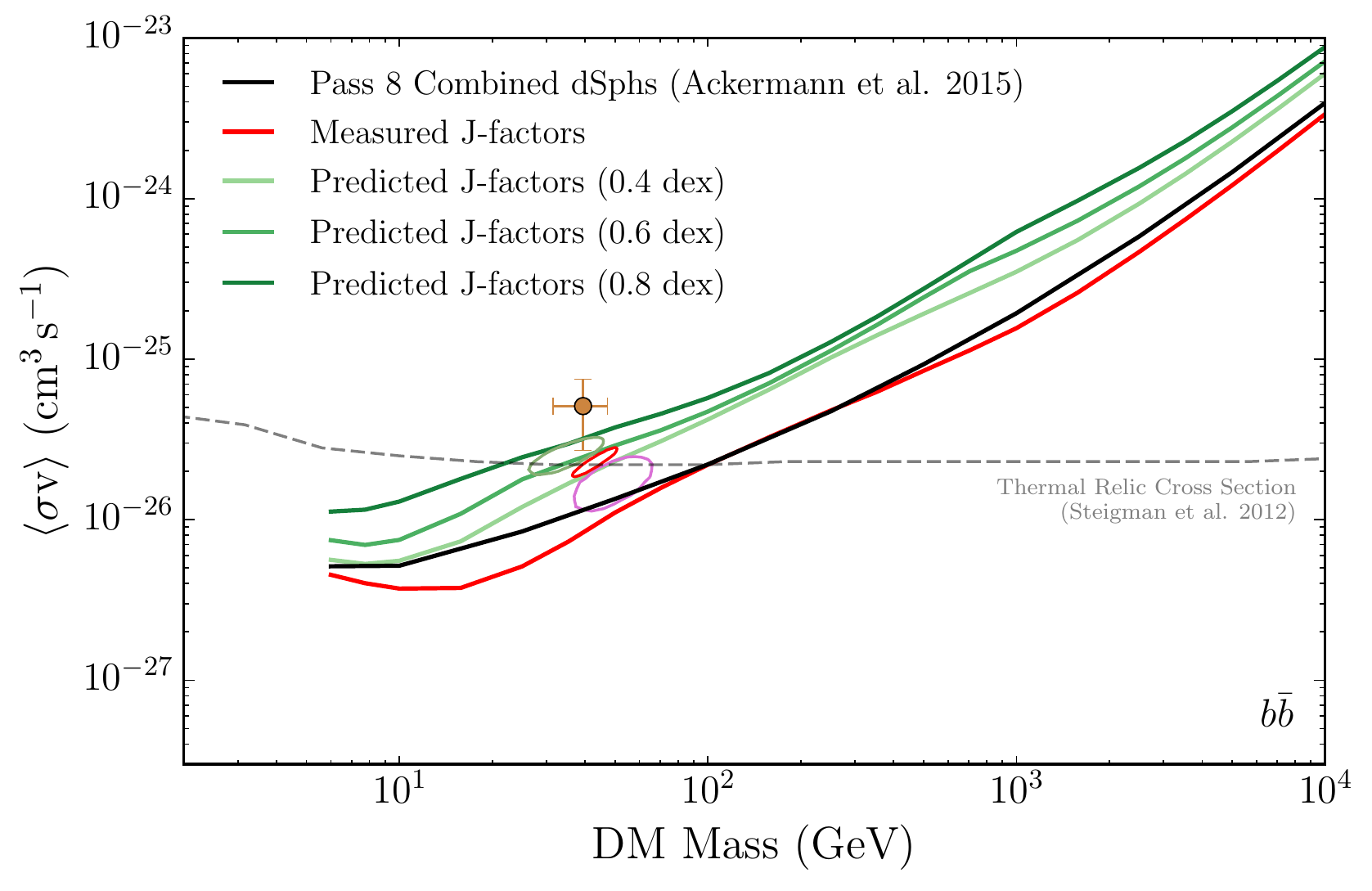} 
  \includegraphics[width=0.49\textwidth]{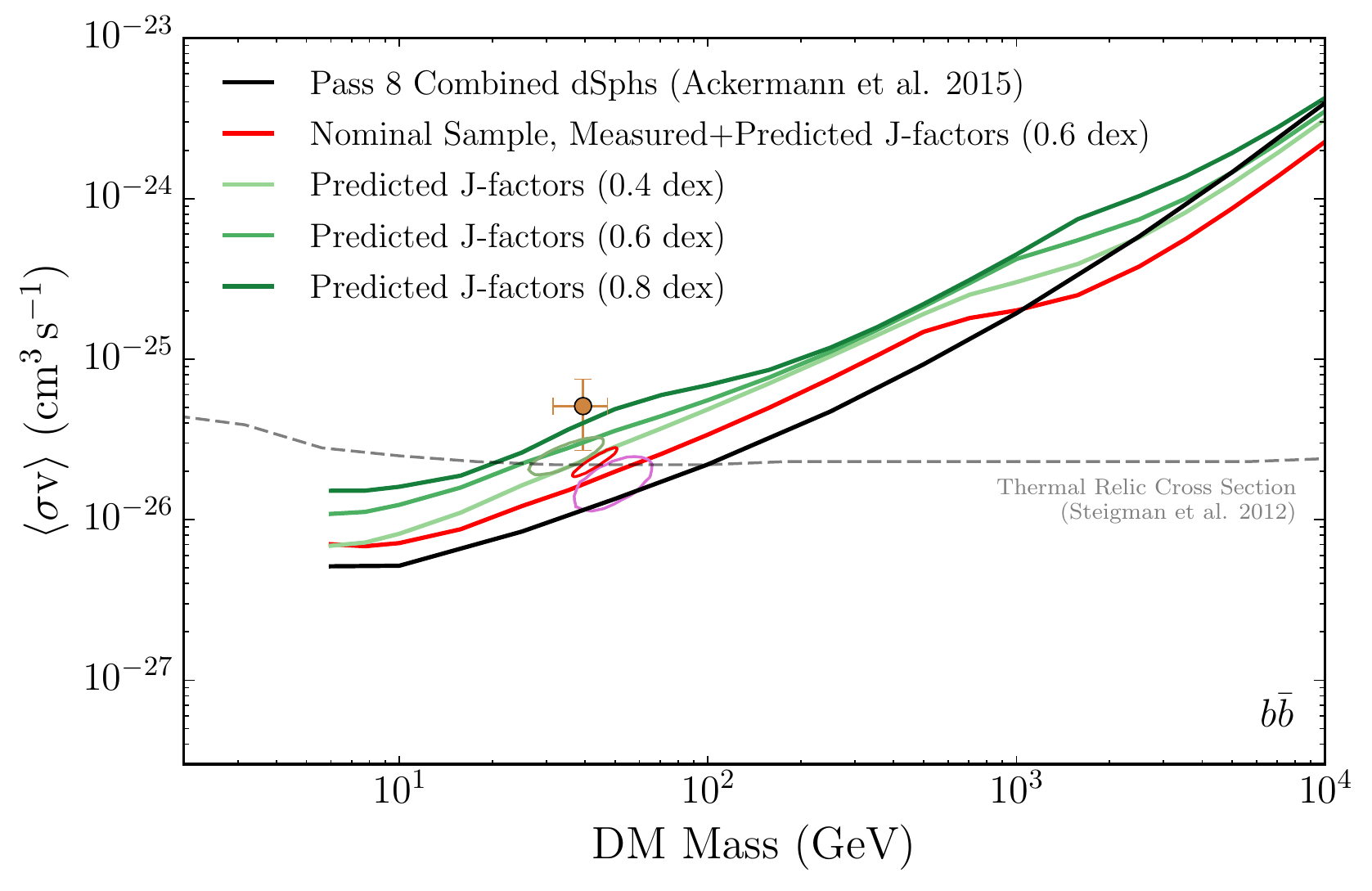} 
  \caption{Upper limits on the DM annihilation cross section (\bbbar
  channel) derived from the sub-sample of dSphs with
  measured \Jfactors (\textit{left}) and the complete nominal sample
  (\textit{right}).  Green curves show the limits obtained when these
  samples are analyzed using only predicted \Jfactors (even when
  measured \Jfactors are available) and fixed \Jfactor uncertainties
  of 0.4, 0.6, and 0.8 dex.  The solid black line shows the observed
  limit from \citet{Ackermann:2015zua}. The closed contours and marker are 
      the same as depicted in \figref{expectUL} and \figref{resultsUL}.}
  \label{fig:pred_jfactor_limits}
\end{figure*}

To characterize the influence on the observed limits due to targets lacking measured \Jfactors, we perform an {\it a posteriori} examination of different subsets of the target sample.
Analyzing only the dSphs with measured \Jfactors yields
results that are similar to those of \citet{Ackermann:2015zua} because
the target sample is nearly identical
(\figref{pred_jfactor_limits}). The primary differences between these
two analyses are the method for calculating \Jfactors and their uncertainties (\citealt{Martinez:2013els}
vs.\ \citealt{Geringer-Sameth:2014qqa}), the spatial model of the dSphs (spatially extended NFW vs.\ point-like), and the addition of the dSph Reticulum~II. 
As shown in the left panel of
\figref{pred_jfactor_limits}, the limits derived with the subset of
dSphs in the nominal sample that have measured \Jfactors deviate by at
most a factor of two from the limits of \citet{Ackermann:2015zua}.

Replacing the measured \Jfactors of all dSphs with the
predicted \Jfactors from \tabref{targets} has a somewhat larger effect
on the observed limits than using only dSphs with measured \Jfactors.  The green
curves in \figref{pred_jfactor_limits} show the composite limits
evaluated using only predicted \Jfactors for two different target
samples: the subset of the nominal sample with measured \Jfactors (left panel) and the full nominal sample (right panel).  
Using only predicted \Jfactors  weakens the observed limits by a factor of 2--3
depending on the choice of \Jfactor uncertainty.

The weakening of the limits when using predicted \Jfactors can be
partially attributed to statistical fluctuations that occur when
changing the relative weighting of objects in the sample.  As a
population, the predicted \Jfactors are statistically consistent with
the measured \Jfactors of \citet[][see \figref{jfactors}]{Geringer-Sameth:2014qqa}.  However, on an object-by-object basis, the differences in \Jfactors can
substantially change the weight given to positive or negative
residuals observed at the location of a given dSph.  We also note that the
limits evaluated with predicted \Jfactors are consistent with the 95\%
expectation band from \figref{resultsUL} (\ie, the expected
variation from statistical fluctuations).

\subsection{Effect of Variations in Diffuse Background Intensity}

\begin{figure*}[h]
  \centering
  \includegraphics[width=0.49\textwidth]{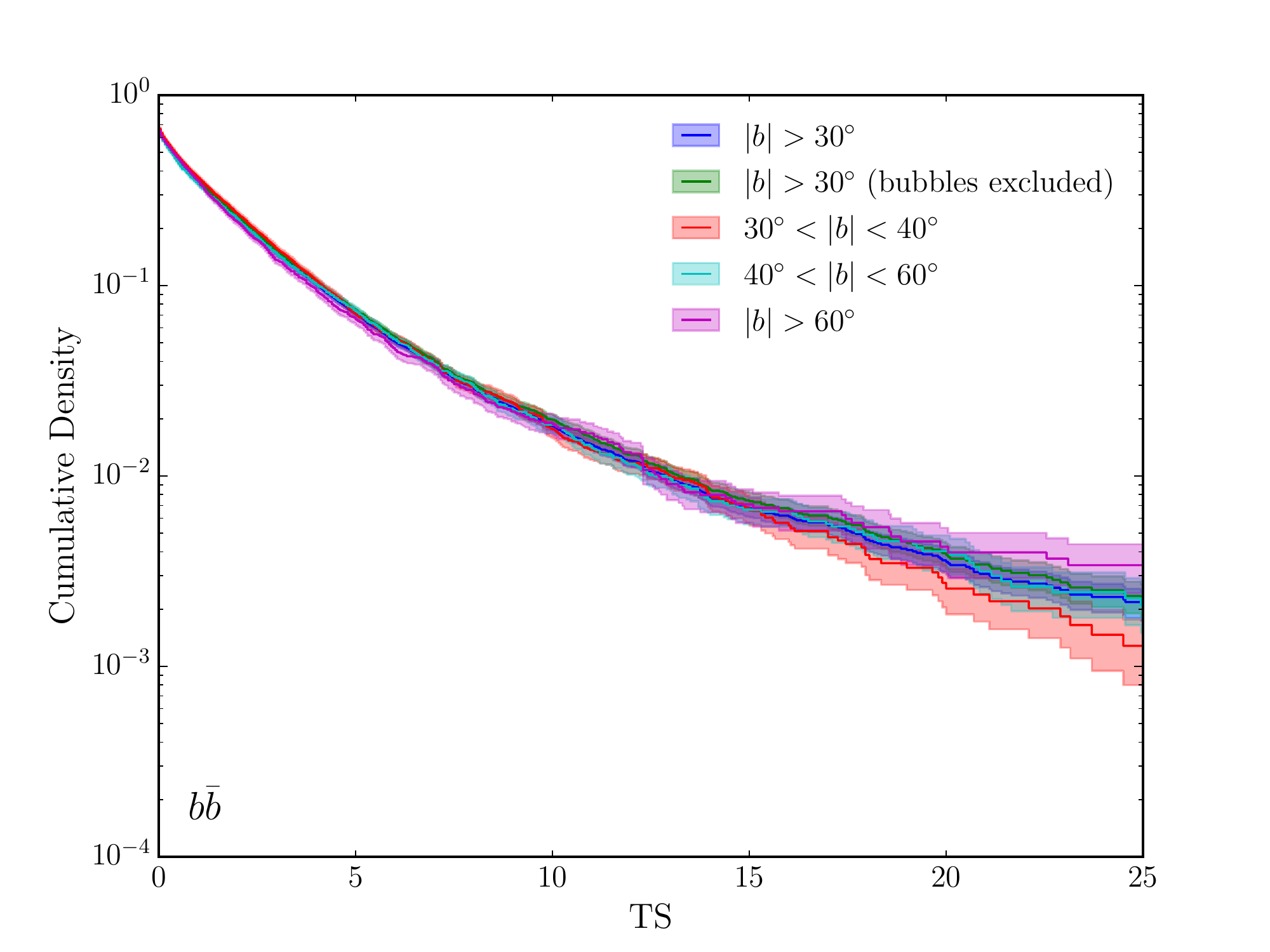}
  \includegraphics[width=0.49\textwidth]{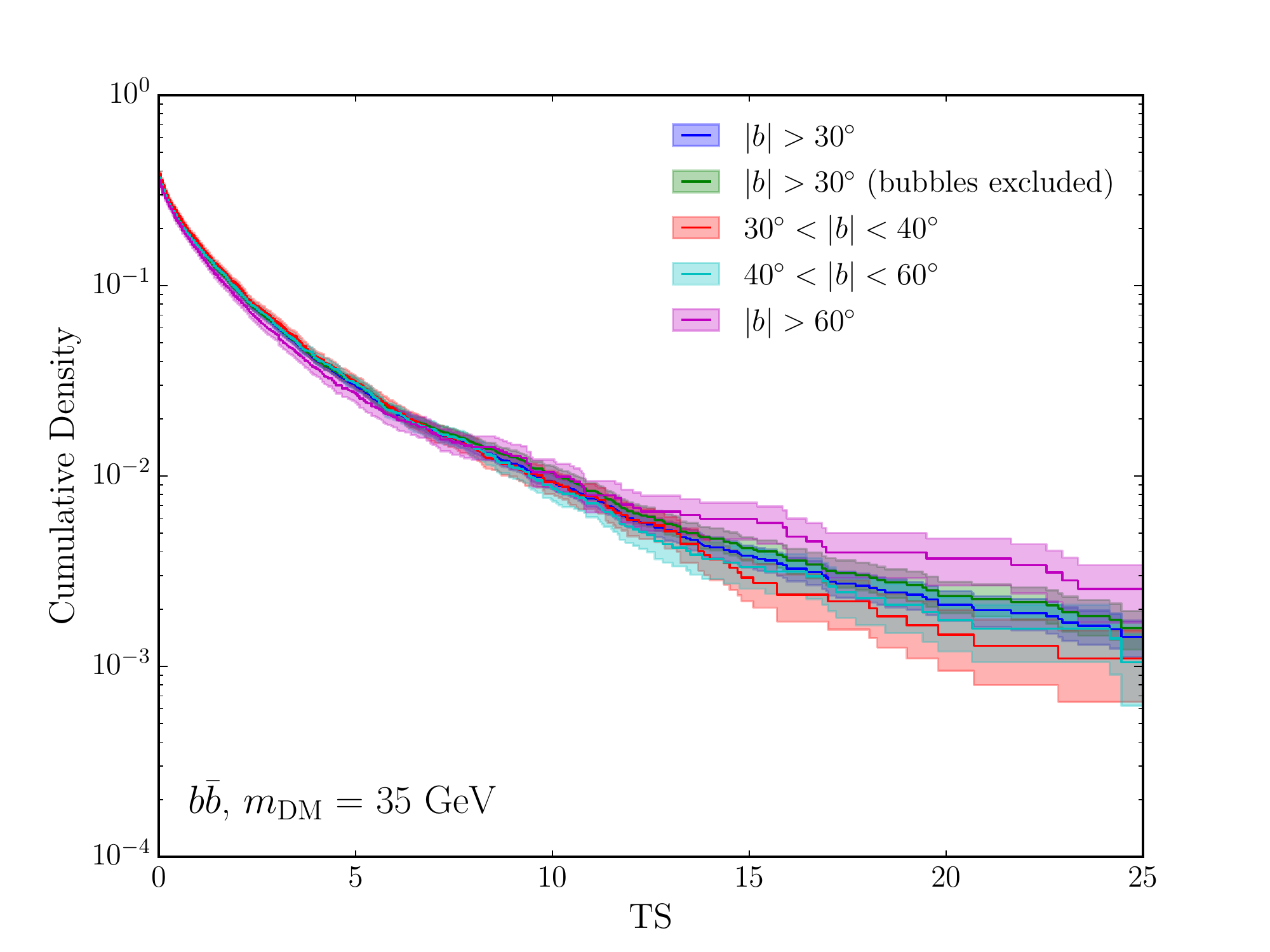}
  \includegraphics[width=0.49\textwidth]{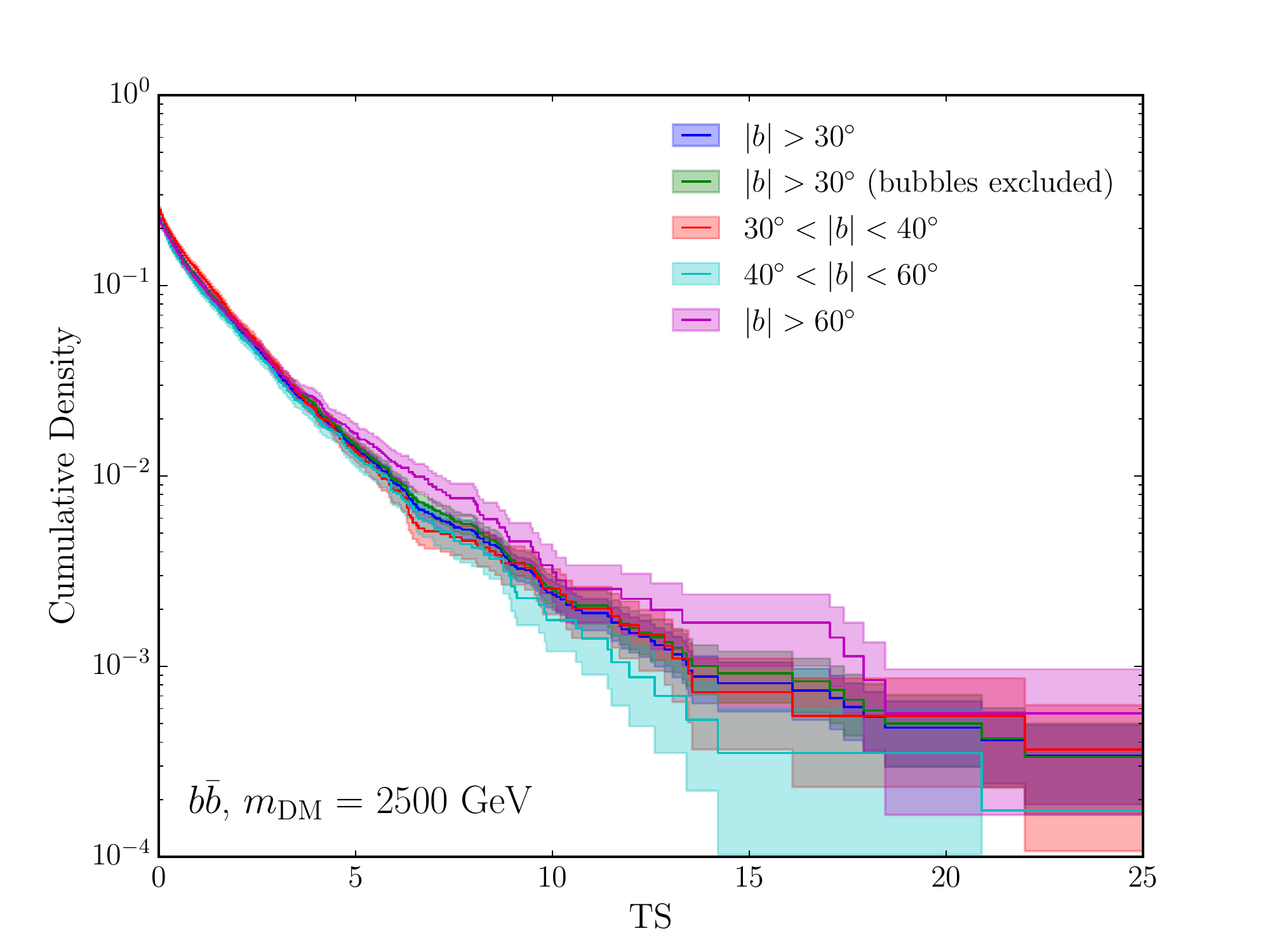}
  \caption{Cumulative TS distributions derived from blank-sky regions
    when fit with DM annihilation spectra ($b\bar{b}$ channel).  Each
    curve corresponds to the distribution of blank-sky positions drawn
    from different subselections of the $|b| > 30\degree$ region.  The
    \textit{bubbles excluded} selection excludes the region containing
    the \textit{Fermi} bubbles with $|\ell| < 30\degree$ and
    $|b| < 60\degree$.  Shaded bands indicate the $1\sigma$
    uncertainties on the cumulative fraction.  \textit{Top Left:} TS
    distribution for the best-fit (maximum TS) mass for each blank-sky
    position and realization.  \textit{Top Right:} TS distribution
    for a DM mass of 35.4\GeV. \textit{Bottom:} TS distribution for a
    DM mass of 2.5\TeV.}\label{fig:blanksky_ts_hist}
\end{figure*}

To assess the potential impact of variations in the diffuse
background intensity on the derived \TS distribution, we have
generated \TS distributions from different subselections of the
blank-sky regions.  As shown in \figref{blanksky_ts_hist}, there is no
statistically significant change in the \TS distribution when
excluding regions around the Fermi bubbles or selecting random sky
positions from lower or higher Galactic latitudes.  We conclude that
the trials factor calculation is robust to local variations in the
background intensity at the level found in the high-latitude sky.

\end{document}